\documentclass[preprint,12pt]{elsarticle}
\usepackage{amssymb}
\usepackage{xspace}
\newcommand{\name}{SemDiff\xspace}
\newcommand{\mypara}[1]{\vspace{2pt}\noindent\textbf{{#1: }}}
\newcommand{\eat}[1]{}
\newcommand{\tool}[1]{\mbox{\small{\textit{#1}}}}
\usepackage{xurl}
\newcommand{\func}[1]{\mbox{\small{\texttt{#1}}}}
\usepackage{hyperref}
\usepackage{algorithmic}
\usepackage{array}
\usepackage[linesnumbered,ruled]{algorithm2e}
\usepackage{multirow}
\usepackage{bbding}
\usepackage{amsmath}
\usepackage{enumitem}
\usepackage{rotating}
\usepackage{xcolor}
\usepackage{amssymb}
\usepackage{amsmath}
\usepackage[table]{xcolor}

\journal{}

\begin{document}

\begin{frontmatter}



\title{\name: Binary Similarity Detection by Diffing Key-Semantics Graphs}


%\author{} %% Author name

%\affiliation{organization={},%Department and Organization

\author[1,3]{Zian Liu
}

\author[1]{Xiaoxi Meng
}

\author[1]{Viet Vo
}

\author[4]{Siqi Ma
}      

\author[2]{Chao Chen
}

\author[3]{Dongxi Liu
}

\author[1]{Jun Zhang
}

\author[3]{Shigang Liu
}

\author[3]{Muhammad Ejaz Ahmed
}

\author[1]{Yang Xiang
}

\affiliation[1]{organization={Swinburne University of Technology},%Department and Organization
            addressline={1 John Street}, 
            city={Melbourne},
            postcode={3122}, 
            state={VIC},
            country={Australia}}  
\affiliation[2]{organization={Royal Melbourne Institution of Technology},%Department and Organization
            addressline={124 La Trobe Street}, 
            city={Melbourne},
            postcode={3000}, 
            state={VIC},
            country={Australia}} 
\affiliation[3]{organization={Data 61, CSIRO},%Department and Organization
            addressline={Corner Vimiera \&, Pembroke Rd}, 
            city={Sydney},
            postcode={2122}, 
            state={NSW},
            country={Australia}} 

\affiliation[4]{organization={University of New South Wales},%Department and Organization
addressline={Northcott Dr}, 
city={Canberra},
postcode={2612}, 
state={ACT},
country={Australia}}

%% Abstract
\begin{abstract}
Binary similarity detection is a critical technique 
that has been applied in many real-world scenarios where source code is not available, e.g., bug search, malware analysis, and code plagiarism detection.
Existing works are \emph{ineffective} in detecting similar binaries in cases where different compiling optimizations, compilers, source code versions, or obfuscation are deployed.

We observe that all the cases do not change a binary's key code behaviors although they significantly modify its syntax and structure. With this key observation, we extract a set of \emph{key} instructions from a binary to capture its key code behaviors. By detecting the similarity between two binaries' key instructions, we can address well the ineffectiveness limitation of existing works. 
Specifically, we translate each extracted key instruction into a self-defined key expression, generating a key-semantics graph based on the binary's control flow. Each node in the key-semantics graph denotes a key instruction, and the node attribute is the key expression. To quantify the similarity between two given key-semantics graphs, we first serialize each graph into a sequence of key expressions by topological sort. Then, we tokenize and concatenate key expressions to generate token lists. We calculate the locality-sensitive hash value for all token lists and quantify their similarity. %We implement a prototype, called \name, consisting of two modules: graph generation and graph diffing. The first module generates a pair of key-semantics graphs and the second module diffs the graphs. 
Our evaluation results show that overall, \name outperforms state-of-the-art tools when detecting the similarity of binaries generated from different optimization levels, compilers, and obfuscations. \name is also effective for library version search and finding similar vulnerabilities in firmware.
\end{abstract}

\begin{graphicalabstract}
\end{graphicalabstract}

\begin{highlights}
\item Research highlight 1: We proposed a novel approach to detect binary code similarity semantically. We implemented this approach through key instructions only and symbolic execution. We translated the instruction summary into a graph in our approach. We proposed an LSH-based
approach and graph serializing to achieve an accurate comparison.
\item Research highlight 2: SemDiff outperforms state-of-the-art tools on 2,106 binaries. Semdiff can also assist the other tools that require
binary comparison.
\end{highlights}

%% Keywords
\begin{keyword}



\end{keyword}

\end{frontmatter}

\section{Introduction}
\label{sec:intro}

Binary code similarity detection (also known as binary diffing) is important for bug search, patch generation and analysis, malware detection, and plagiarism detection~\cite{binarysurvey}. Existing works can be categorized into machine-learning-based approaches or program-analysis-based approaches.

\noindent
\textbf{Machine learning based approaches} either capture syntactic, structural, and semantic features from binary code to train a similarity detection model\textcolor{black}{~\cite{Gemini,aDiff,VULSEEKER,genius,ordermatters,graph-cross,interpretable,asteria}} or leverage natural language processing (NLP)\textcolor{black}{~\cite{asm2vec,deepbindiff,aDiff,innereye,Kam1n0,jtrans,Palmtree,uniasm,safe,jason}} to learn semantic information.
Although such machine-learning-based approaches are theoretically effective, 
    the performance highly relies on how well the training data is created.
%However, the labor works of collecting training data (i.e., summarizing syntactic, structural, and semantic features) are time-consuming. 
%In addition, 
    Real-world binaries are diverse, that is, the same piece of source code could be compiled into different pieces of binaries because of using different compilers (optimization levels)~\cite{deepbindiff}.
It is thus difficult to build a representative dataset for training.
If the training data are not well established, 
    the corresponding approaches will be affected significantly. Unsupervised learning automatically learns each instruction embeddings from its context instructions. However, optimizations or different compilers can make changes to a non-trivial part of the function, rendering this method less accurate. On the other hand, machine learning-based methods aim to preserve semantic information by predicting assembly syntactic embeddings, though these embeddings may not accurately retain the exact semantics.

\noindent
\textbf{Program analysis approaches} generally execute static/dynamic analysis to extract information (e.g., data/control dependencies) from the binaries and then quantify similarity based on certain defined rules\textcolor{black}{~\cite{binsim, IMF-SIM, COP,SMIT,BINCLONE,SPAIN,IDEA,MBC,Expose,exediff,tracy,binsequence,binslayer,BEAGLE,fossil,SIGMA}}. 
However, 
    static function-level processing approaches match a sequence of blocks or instructions, 
    which defines each basic block or each instruction within a function as the smallest unit of similarity quantification. 
Considering the case of using different compiling optimization to proceed with the same source code, 
    the basic block of each function will become totally different through block splitting, instructions can be different due to instruction substitution.
Therefore, 
    block or instruction comparison is not the ideal approach to handle binary similarity comparison\cite{unleashing}. 
%Although dynamic analysis could resolve the program of binary code diversity through runtime program execution, 
%    it cannot achieve a full code coverage~\cite{binsim,binarysurvey}. 
    
To resolve the above limitations, 
    we propose and implement a novel semantic-aware approach, 
    \name, 
    to compare arbitrary binary codes without considering which compilers and optimization levels the developers utilized.
Specifically, 
    \name consists of two modules, \emph{Graph Generation} and \emph{Graph Diffing}.
In graph generation,
    \name takes as input of a pair of binary functions and then constructs a key-semantics graph for each function by 1) identifying the key instructions that reflect major function behaviors (e.g., invoking functions, assigning value); 2) applying symbolic execution to extract symbolic expressions of each instruction and instruction dependencies; 3) translating symbolic expressions into self-defined key expressions and using directed edges to connect all the correlated instructions.
After having a pair of key-semantic graphs of the two binary functions, 
    graph diffing takes the graphs as input and leverages topological sort to serialize both graphs into two sequences.
It then tokenizes each sequence and executes locality-sensitive hash (LSH)-based comparison to calculate an LSH value for each key-semantics graph.
Two graphs (i.e., binary functions) are regarded as similar if their LSH values are similar. %\sq{check if this is correct}.
%\sq{add more technical steps if needed}

\eat{
we perform a static analysis of a binary through symbolic execution and extract only key instructions containing opcodes and operands. Then 
we translate each key instruction into an Intermediate Representation (IR) that is defined by ourselves. Then we generate a graph of IRs. Each node represents a key instruction in this graph, and its attribute is an IR.
Essentially, binary diffing or binary similarity detection has been converted into attributed graph diffing or similarity detection, which is, however, still challenging because we cannot directly use existing techniques~\cite{asm2vec,deepbindiff} to diff a pair of attributed graphs. Moreover, comparing two attributed graphs directly is time-consuming because one needs to consider each node along with its neighbors' relations.
Thus in \emph{graph diffing}, to simplify graph diffing, we first transform a pair of attributed graphs into two sequences by \emph{topological sort}. Similar binaries with similar control flows can result in similar topological sequences. We then diff the sequential attributed graphs by \emph{locality-sensitive hash}. Since similar sequences result in similar LSH hash values.
}

\textcolor{black}{We evaluated the binary similarity performance of \name using nine libraries: openssl, libtomcrypt, coreutils, ImageMagick, libgmp, curl, sqlite3, zlib and Puttygen. We compared \name with five state-of-the-art tools: Asm2Vec~\cite{asm2vec}, Gemini~\cite{Gemini}, Palmtree~\cite{Palmtree}, JTrans~\cite{jtrans}, and UniASM~\cite{uniasm}.}
The results demonstrated that \name on average outperforms baseline tools no matter whether the source code is compiled by the same compiler (with different optimization levels) or the same optimization levels (with different compilers). 
%\sq{what's the reason for the outperformance?}
The out-performance of \name is because of the semantic-preserving key expressions and effective LSH-based graph diffing.
In addition, 
    we applied \name to conduct vulnerability detection and library version check and found that \name could significantly improve the performance of such similarity-dependent detection.

%in their effectiveness in detecting similarity of binaries from different compiling optimization levels, compilers and source-code program versions. Overall, \name outperforms the five tools in detection effectiveness. 
%Also, {we choose two representative machine-learning-based tools (i.e., Gemini and Palmtree) and compare them with \name in obfuscated binary similarity detection
%Our experimental results clearly show that \name performs better than Gemini~\cite{Gemini} and Palmtree~\cite{Palmtree} in all three obfuscation options (i.e., SUB, FLA, and BCF) provided by OLLVM~\cite{ollvm}.}
%Our current version of \name works for x86-based binaries, and we will discuss how to extend it to support cross-architecture similarity detection in our future work in \autoref{sec:conclusion}.
%All the binary code are the same architecture. Currently we only support x86 binaries since they are most prevalent worldwide. Our method can be extended to other single-architecture comparison. It is also possible to compare cross-architecture binary code since we are using the Intermediate Representation (IR) in our method. We leave it in future work.
%including a widely used commercial tool Bindiff~\cite{bindiff}, an open source binary diffing tool released by Google called functionsimsearch~\cite{functionsimsearch}, Asm2Vec~\cite{asm2vec}, and Kam1n0\cite{Kam1n0}.

\vspace{0.1cm}
\noindent
\textbf{Summary of Contributions:} 

\begin{table}[h!]
\centering
\begin{tabular}{|c|c|p{3cm}|p{2cm}|}
\hline
\textbf{Method} & \textbf{Approaches} & \textbf{Assembly syntactic Embedding} & \textbf{Training required}\\
\hline
%Bindiff \cite{bindiff} & Complex rule & \Checkmark &\XSolidBrush\\
Asm2Vec \cite{asm2vec} & Unsupervised neural network & \Checkmark &\Checkmark\\
Gemini \cite{Gemini} & Supervised neural network & \Checkmark & \Checkmark \\
Palmtree \cite{Palmtree} &Supervised neural network&\Checkmark &\Checkmark\\
JTrans \cite{jtrans}&Unsupervised neural network&\Checkmark &\Checkmark\\
UniASM \cite{uniasm}&Unsupervised neural network&\Checkmark &\Checkmark\\
\name (Ours)&Simple rule&\XSolidBrush&\XSolidBrush\\
\hline
\end{tabular}
\caption{Comparison of our work with others}
\label{tab:method_compare}
\end{table}

We summarize our major contributions as follows:
\begin{itemize}[noitemsep, topsep=2pt, partopsep=0pt,leftmargin=0.4cm]
\item We proposed a novel semantic-aware approach for binary similarity detection. 
    We abstract binary code by selecting the key instructions only and then executing symbolic execution to extract the instruction correlations for further analysis. 
    %translate and simplify all binary code. 
The approach efficiently simplifies binary code by refining exact semantic information.

\item We proposed an approach to translate the instruction summary into a graph. 
To achieve an accurate comparison, we propose an LSH-based approach to convert a graph into a sequence for the final similarity calculation.
Such an approach could be applied to various scenarios (e.g., vulnerability searching, and malware detection). The comparison rules are very simple and our method does not require training.

%\item We observe that different compiling optimization, compilers, source-code program versions, and obfuscation options do not change the key
%code behaviors of a binary. With this key observation, we extract key instructions from a binary and transform them into a key-semantics graph. We then combine {topological} sort and locality-sensitive hash to compute the similarity between two given key-semantics graphs. The source code and data are available at \\ \href{https://github.com/SemDiff4BinaryDetection/SemDiff4BinaryDetection}{https://github.com/SemDiff4BinaryDetection/SemDiff4BinaryDetection}.

\item We assessed the performance of \name by using 2,106 binaries (considering various compilers, compiling options, source code versions, and obfuscations) in the 52 packages. and also compared the detection results with the state-of-the-art tools. 
The results demonstrated that \name not only outperformed in binary similarity detection but also can be utilized to assist the other tools that require binary comparison. \textcolor{black}{The results show that \name achieves higher accuracy than GNN-based methods.}   

\item We currently published our tool, \name, and the experimental data to our repository \url{https://anonymous.4open.science/r/SemDiff4BinaryDetection-F12C/README.md} for reviewers to check and test. After this paper is accepted, the tool and dataset will be published. 

\end{itemize}

Note that binary code in this paper represents the assembly code compiled by compilers.

\begin{figure}[h]
\centering
\includegraphics[width=0.4\textwidth]{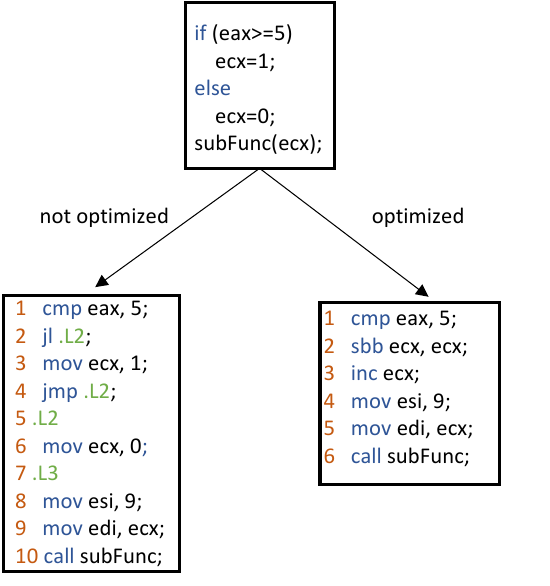}
\caption{Due to compiling optimization, a binary snippet on the left is obviously different in syntax and code structure from the one on the right, although they are from the same source code (on the top).
}
\label{fig:optimization_example}
%\vspace{-0.3cm}
\end{figure}

\section{Background}
\label{sec:background}

\subsection{Motivation}
%\sq{example here}
We demonstrate a code compilation example in \autoref{fig:optimization_example} to represent the binary diversity and how a compiling optimization changes a control-flow-graph (CFG). Specifically, a non-optimized binary snippet (on the left) has a conditional branch (\texttt{jle .L2}) with two destinations, i.e., \texttt{mov ecx, 1} and \texttt{mov ecx, 0}. An optimized snippet (on the right) substitutes the three instructions with \texttt{sbb}. Although both snippets are from the same source code, they have different syntax and structures. The performance of supervised learning heavily depends on the training data. Therefore, supervised learning approaches \cite{Kam1n0,Gemini, genius, VULSEEKER} will be ineffective for binaries compiled with unseen optimizations or compilers. Unsupervised learning automatically learns instruction embeddings from contexts. Therefore, if a non-trivial part of the context is different due to optimizations or different compilers, this approach \cite{deepbindiff,asm2vec} will be less accurate. Some program-analysis-based methods compare similarity at the granularity of basic blocks. However, as in the example, the blocks are merged after optimization. Thus methods based on basic-block level \cite{COP, bingo,bindiff} comparison can be ineffective for optimizations. Some program-analysis-based methods utilize Longest Common Sequence (LCS) to align two sequences of instructions or blocks from the two functions. However, optimization as shown in the example can change the instruction or basic block or their ordering, thus posing challenges to sequence aligning-based approaches \cite{binsim, IMF-SIM, COP}.

\subsection{Preliminary}

In this section, we first present our goal and then give our assumptions of \name. 
%We then describe an overview of the \name prototype.
%In this section, we first define the problem of binary code similarity detection. Then we describe the overview of the methodology.

\mypara{Our goal}
Detecting binary similarity can be conducted at different granularity, i.e., basic block, function, or inter-procedure CFG. A basic-block level quantifies the similarity between two given basic blocks. A function level quantifies the similarity between two given functions. An inter-procedure CFG quantifies two graphs of basic blocks {connected across multiple functions} following the control flow of a binary. %An inter-procedure-CFG level quantifies the similarity between two given inter-procedure CFGs. %This is similar to appending subfunctions at the function level. 
Given that it is critical to find vulnerabilities in similar functions in the real world~\cite{VULSEEKER, bingo, genius}, our goal is to effectively detect binary similarity at a function level, the same as ~\cite{asm2vec,bindiff,functionsimsearch, Kam1n0, COP}. %There are multiple commercial tools (e.g., \texttt{IDA Pro} in our work) that can successfully reverse-engineer a binary into a control-flow graph of assembly functions.
As mentioned in ~\autoref{sec:intro}, binary similarity detection includes machine-learning and program-analysis approaches. In this paper, we use the program-analysis approach to compare binary similarity.

%\mypara{Definition of Semantic Similarity}
%\Shigang{use symbols such as function A and Function B are semantic similar to each other given that xxx }
\mypara{Our Assumptions}

Similar to previous works~\cite{asm2vec,deepbindiff,innereye,binsim}, we make three assumptions that are practical in the real world. First, binaries are stripped without debugging information, as a stripped binary is released as a software product in the real world to protect its intellectual property. 
Second, {binaries can be obfuscated, which is often used to make real binary code difficult to understand.}
Third, binaries are assumed to be unpacked, as binary unpacking is an orthogonal problem and can be solved by prior works~\cite{cheng2018towards,roundy2013binary}.

\subsection{Challenge}
To compute the similarity of the two binaries precisely,
    the following challenges need to be resolved.

%\vspace{0.1cm}
\noindent
\textbf{Challenge I: Extract Core Semantics.}
The syntax of the binary code varies when compiling the same source code with different compilers or optimizations. Extracting equivalent semantics from those two syntactically different programs is challenging. Existing solutions typically compare symbolic expressions after translating the binary into higher-level Intermediate Representations (IR) \cite{bingo,  COP}. However, these IRs do not simplify the binary code. On the contrary, their grammar makes IR even more complex than binary code because they tend to translate one binary instruction to multiple IR instructions. Moreover, even translating to IR, IR still contains unmatched variables due to different compilers or optimizations. Therefore, comparing symbolic expressions of IR still cannot accurately identify semantic equivalence. Also, each binary instruction corresponds to at least one IR instruction. Therefore, a representation of the binary code that can both preserve semantics and simplify binary code is required. 
\noindent
\textbf{Challenge II: Symbolic Formula Too Long.}
Instead of processing simple text-based analysis, 
    many approaches~\cite{asm2vec,VULSEEKER,innereye,Gemini} study the code similarities through semantic graph comparison.
They extract graphs (e.g., control flow graphs, call graphs) and convert the graphs into vectors.
Through machine learning algorithms, 
    the binary code snippets with similar vectors will be identified. 
However, 
    these approaches are not applicable.
%to our defined key expressions. 
Specifically,
    those approaches regard each binary instruction as a vector of a fixed size, as different binary instruction generally are short (e.g., one operator with one or two operands) and has similar lengths. %binary code using only one statement to express a function/variable execution, 
     %binary code might use multiple lines of instructions (e.g., \func{mov}, \func{add}) to complete the execution. 
In reality, such a fixed size requirement is not applicable to our case, as our designed key expression might contain the semantics of multiple lines of instructions, thus can have extremely long expression, and the expression length can vary enormously. 
%In reality, such a fixed size requirement is not scalable and it is difficult to determine a fixed size that could be applied to all scenarios.
%     However, our designed key expression might contain the semantics of multiple lines of instructions, thus can have extremely long expression and the expression length can vary enormously. Therefore, it is challenging to represent each key expression as a vector.
%Without matching the instructions with each source code statement,
 %   it is difficult to convert graphs of binary code into vectors.
Instead,
    we adopt a locality-Sensitive hash (LSH) based approach to compute a hash value for all the nodes in the graph and further compare the hash values to calculate the similarity. LSH algorithm is able to represent high-dimensional data with low-dimensional data while preserving the relative distance among data.
%The key-semantics graph consists of multiple nodes. Each node is a key instruction and its attribute is its key expression, the symbolic expression we extracted. To compare the similarities of instructions, machine-learning based methods in \cite{asm2vec,VULSEEKER,innereye,Gemini} are not applicable here because they transform each binary instruction into a vector to represent the semantics. Unlike the binary instructions which are generally similar in length, the key expressions can be extremely complex and varies significantly in length. %Therefore it is hard to directly use their methods. We topologically order the key expressions and hash them to output a function-level semantic hash value and compare the hash value's similarity.

%\vspace{0.1cm}
\noindent
\textbf{Challenge III: Inefficient and Inaccurate Sequence Comparison}
%When comparing the similarity of two graphs, 
    %most of the existing approaches~\cite{COP,IMF-SIM} leverage \tool{Theorem Prover} and \tool{longest common sequence (LCS)}. \Zian{Therom Prover is widely used to decide symbolic formula equivalence. It can identify semantically equivalent but syntactically different expressions. But Therom Prover is very expensive \cite{COP}.}
    %\sq{how does theorem prover work?}.
To compare the function similarity, many works adopt a divide-and-conquer algorithm: longest common sequence (LCS) \cite{binsim,IMF-SIM,COP}. In LCS, each function is treated as a sequence of instructions or blocks. At the lowest level, it compares one instruction with another instruction or one block with another block. However, the comparison strategy at the lowest level is to be defined by us (i.e., how to compare the similarity between two instructions or two blocks). Therefore, human experts need to inspect the binary code and conclude rules for comparison. The accuracy heavily depends on the quality of the set comparison rules. Also, it is time-consuming for human to propose rules, making this approach unscalable. Moreover, LCS is NP-hard, the running time cost is significant. %\sq{what does "compare blocks accumulatively" mean?}.
Nonetheless, 
    these approaches cannot process the graph of binary codes efficiently.
%\sq{why do these approaches cannot be applied, except for the overhead problem?}
    
%To compare the similarities of instructions, program-analysis based methods like \cite{COP,IMF-SIM} use Theorem Prover or longest-common-sequence (LCS) methods to compare the symbolic equivalence. Theorem Prover is known to to have computationally high overhead. We implemented the LCS-based algorithm but it both take long running time and low accuracy because LCS algorithm compares each block accumulatively, rather than as a whole on function-level. %Therefore, we use topological sort and hashing to solve this problem.

\subsection{Solution}

%In this section, We firstly define key-semantics graph. Then we define key instructions and key expressions.
%\subsubsection{Key-semantics graph}
Regarding the challenges to binary comparison, 
    we propose the following approaches.
    
Respond to challenge 1, we propose a symbolic-based binary translation approach to abstract key instructions from each binary and generate semantic-aware representatives for further comparison.
After observing the binary codes generated from the same high-level source code, 
    we found that instructions can be classified into \emph{key instructions} and \emph{non-key instructions}. 
In particular, 
    key instructions represent the major function executions and parameter value transmission and non-key instructions are the instructions for preprocessing purposes such as address computation.
    
Through our manual inspection, 
    we classified the key instructions into four types, 
    \emph{calling behavior}, \emph{comparing manner}, \emph{indirect branch}, and \emph{memory store}.

\begin{itemize}[noitemsep, topsep=2pt, partopsep=0pt,leftmargin=0.4cm]
\item \textbf{Calling behavior.} It represents a calling instruction that takes operands operated by previous instructions as function arguments.
%For example, \texttt{call eax} is an indirect call and the value of \texttt{eax} is determined by previous instructions. Also, \texttt{mov edi, 5} followed by \texttt{call subFunc} means that \texttt{subFunc} is called with an input parameter of 5.
%can have input parameters generated by previous instructions, e.g., Also, a calling convention can have operands as target addresses, e.g.,. 
%We note that \texttt{call subFunc} might carry no parameter and we still regard it as a calling behavior.
%We still use this category of instruction as carrying no parameter is also a comparable feature. %(e.g.,  \texttt{call subFunc}) 

\item \textbf{Comparing manner.} It is an instruction with operands of comparing objectives.
The instruction with the operator such as \func{cmp}, \func{test} %\func{sub}, \func{and}, \func{or} 
will affect which subsequent branches to execute at a joint point.

\item \textbf{Indirect branch.}
It represents an instruction with an operand of a target address (e.g., \func{jmp eax}). 
%as it provides sufficient semantics in symbolic expression of \texttt{eax}. 
%\sq{The following sentence is confusing - it says the direct branch is not a key instruction, then why do we list it here?}
%We note that a direct branch is not a key instruction regardless of whether it is conditional (e.g., \texttt{jnz address}) or unconditional (e.g., \texttt{jmp address}) as their operands are not computed from previous non-key instructions.
%because it provides no semantic gain other than implying it is a branching instruction. This is because the branching destination address is decided at compiling time thus does not has semantic information. Moreover, for the conditional direct branching instruction , 
%\zhi{the two-path branching semantics is reserved in the IR because IR inherit the control flow from CFG.}
%the Indirect branching instruction   since the address can be written in the form of a symbolic expression.
\item \textbf{Memory store.} It is an instruction with the operand that stores values or memory addresses (e.g., \func{mov [edx], ebx}). \textcolor{black}{}
\end{itemize}

\textcolor{black}{Note that, since we use symbolic execution, in the case of indirect branches, the value of eax will be abstracted as a symbolic expression. For example, at some point in the assembly code, \texttt{eax} might be represented as \texttt{(VAR0 + 1000) \& 2}. The exact symbolic expression depends on how the value was computed in the preceding instructions.}

\textcolor{black}{Additionally, since we use symbolic execution, this example instruction of the memory store operation might be translated to \texttt{[VAR1 - 1000] = (VAR3 * 3) + 2}. This is because the symbolic expression for \texttt{edx} was calculated as \texttt{VAR1 - 1000}, while the symbolic value for \texttt{ebx was computed as (VAR3 * 3) + 2} based on the previous instructions. As a result, the impact of compiler optimizations is mitigated.}

%\autoref{tab:symbolic_expression} shows syntax of these four types of key instructions' symbolic expressions. 

\textcolor{black}{We select these four key instructions because theoretically, these four instruction categories collectively capture the essential program semantics along data-flow (through memory writes, calls), control-flow (through compares and indirect jumps), and inter-procedural dimensions (through calls). Other instruction types contribute to operand computation but do not alter the program’s observable behavior beyond these semantic boundaries. Also, in the experiment, the empirical study shows that those four key instructions covers more than 94\% code coverage on average.}

%Although each binary are abstracted into key instructions and these key instructions are classified based on their semantic meanings, 
%    it is still difficult to retrieve the semantics because instructions may be syntactically different. To minimize the syntactic differences, 
%    we define four types of \emph{key expressions} corresponding to the types of key instructions, shown in Table~\ref{tab:IR}. %Similarly, we have four types of key expressions as shown in \autoref{tab:IR}.
%Considering Figure~\ref{fig:optimization_example} as an example,
%    \sq{describe the figure} \Zian{The total instruction number decreased after optimization. Moreover, the changed instructions might have different symbolic expressions. Therefore, only with symbolic expressions of each instruction still can not mitigate impact caused by different compilers or optimizations. }.

\begin{table}[h!]
\centering
\footnotesize
\begin{tabular}{c|c}
\hline
Key Instruction & Key Expression\\
\hline
calling behavior & $RET\_FuncAddr(exp_1, ..., exp_n)$\\
comparing manner & $exp_1\; cmp\; exp_2$\\
indirect branch & $branch\; exp$\\
memory store& $[exp_1]\;=\;exp_2$\\
\hline
\end{tabular}
\caption{Key instructions and their corresponding key expressions.}
\label{tab:IR}
\end{table}

%\begin{equation}
%\begin{split}
%RETURN\_FunctionAddress(exp\_1, ..., exp\_n)
%\end{split}\end{equation} 
Therefore, we further define four types of \emph{key expressions} corresponding to the types of key instructions, shown in Table~\ref{tab:IR}.
For the key expression of a calling behavior, $RET$ denotes a \texttt{call} instruction. {$FuncAddr$} is the starting address of a function, and $exp_i$ ($i\in \{1, ..., n\}$)
is the symbolic expression of the function arguments. 
%It can be a number, a string, a memory address, a function parameter, or a symbolic expression mixed of them. 
For the key expression of a comparing {manner}, 
$cmp$ denotes a comparing instruction. $exp_1$ and $exp_2$ are two operands used for comparison. For the key expression of an indirect branch, $branch$ denotes a branching instruction and $exp$ refers to the symbolic expression of the branch destination. For the key expression of a memory store, $exp_2$ is written into the memory address denoted by $exp_1$. According to the defined key expressions, 
    we first symbolically execute the binary to derive each instruction's each operand's symbolic expression
    and then translate each marked key instruction into specific key expressions (see Section~\ref{sec:graph_gen} for details).

%\vspace{0.1cm}
%\noindent
%\textbf{Definition 1:}
%\subsubsection{Defining Key Expressions}\label{sec:key_expressions}
%Although key semantics can be retrieved by key instructions, the instructions themselves can differ syntactically for various reasons, e.g., they may come from different optimizations as shown in ~\autoref{fig:optimization_example}. 
%To minimize the syntactic differences, firstly we symbolically execute the function so that each instruction's each operand can have a symbolic expression after the execution. Then we translate the key instructions' symbolic expressions into the so-called \textbf{key expressions} (see  ~\autoref{sec:graph_gen} for more details). 
%Then we translate the synthesized expressions into  .  

%\subsubsection{Key Expression}
%\label{sec:key_instruction_translation}

\eat{
\begin{figure}[h]
\centering
%\centerline{\includegraphics[width=\textwidth]{pictures/methodology.pdf}}
\includegraphics[width=0.5\textwidth]{pictures/IR.pdf}
\caption{Our IR syntax \zhi{loop variable rather than iteration variable}}
\label{fig:IR}
\end{figure}
}

To address challenges 2 to 3, we utilize the LSH algorithm. Specifically, given the abstracted binary with key expressions, 
    we raise an LSH-hash-based comparison to compute the similarity of the two abstracted assemblies.
Specifically, 
    we first build a \emph{key-semantics graph}, which summary the major behavior of each function.
Referring to node correlations demonstrated in the graph, 
    we topologically sort all nodes and use LSH hashing to hash each graph into an LSH value.
By comparing the LSH values of the two graphs, 
    we can finally speculate the similarity between the two graphs.

%\noindent\textbf{Solution 2-3:} Secondly and thirdly, to compare the similarities of instructions with higher efficiency and more accuracy, we topologically sort the nodes in the key graph, concatenate each node's attributes, and use LSH hashing to hash each key graph into a LSH value. Comparing the graph similarity now turns into comparing two LSH value similarity.

%\vspace{0.1cm}
%\noindent
%\textbf{Definition 2:}
%A \textbf{key-semantics graph} preserves key semantics (i.e., key code behaviors) in a function of a binary. 
%In the following, we introduce key instructions and key expressions.

%\subsection{Key-Semantics Graph}\label{sec:graph definition}

%Semantics means that each node in the graph contains key expression as its semantic information. Key means that the graph consists of instructions having critical code behaviors, i.e., key instructions. 

%Specifically, each key-semantics graph $G=(V,E)$ 

%\zhi{what is node attribute here?}

%\subsubsection{Defining Key Instructions}

%\input{symbolic_execution.tex}
%\input{representation.tex}
%\input{graph_comparison.tex}
\section{\name}
%In this section, we first describe an overview of \name and then discuss its methodology in detail.
\begin{figure}
\centering
%\centerline{\includegraphics[width=\textwidth]{methodology.pdf}}
\includegraphics[width=\textwidth]{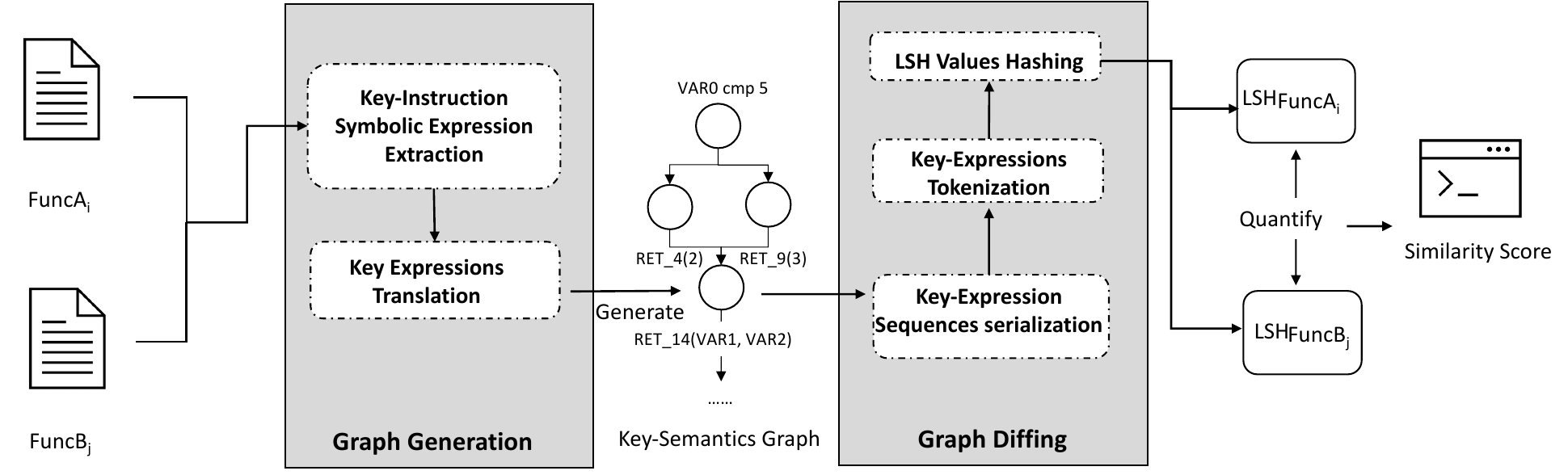}
\caption{\name Overview. \name consists of two modules, i.e., graph generation and graph diffing. (LSH is short for Locality-Sensitive Hash.) %\Shigang{move to the previous page} 
}
\label{fig:design-overview}
\end{figure}
%To compare the similarity of binary code, 
%    we propose and implement a structured and semantic-aware approach, \name.
\subsection{Overview}
%In this section, we give an overview of the \name prototype.
For two binary candidates, e.g., \texttt{BinA} and \texttt{BinB}, each has a set of functions, that is, $\{FuncA_1, ..., FuncA_n\}$ and $\{FuncB_1, ..., FuncB_m\}$.
To detect binary similarity at function level, we select a pair of $FuncA_i$ ($i\in \{1, ..., n\}$) and $FuncB_j$ ($j\in \{1, ..., m\}$) from \texttt{BinA} and \texttt{BinB}, and feed them into \name for similarity quantification. \name consists of two modules, i.e., graph generation and graph diffing, which work as follows:
%As shown in 

%The input fed into \name is a pair of binaries, i.e., . Each binary  

%As shown in \autoref{fig:design-overview}, \name has two modules, 
%In the following sections, An input for \name is a pair of {Binary A} and {Binary B}. For each function in \texttt{Binary A}, we compute its similarity score with each function in \texttt{Binary B} and pick the pair that has the highest score.

\begin{itemize}[noitemsep, topsep=2pt, partopsep=0pt,leftmargin=0.4cm]
\item \textbf{Graph Generation}. It contains three major steps. First, we leverage customized symbolic execution to extract symbolic expressions of key instructions from a given function. 
Second, we translate the extracted symbolic expressions into key expressions.
Last, we generate a graph preserving the key semantics of a function by connecting the translated key expression to the function's control flow. 
%retrieve their semantics by using  %Each operand value of each instruction in the function is then updated as a symbolic expression. 
%We proposed our symbolic traversal methods to cover all instructions for each function. We propose a complete instruction coverage method, and a lightweight processing for loops.
%Only key instructions' symbolic values are of our interest since they also contain the non-key instructions symbolic values. We detect the key instructions based on a set of rules. 
%To get the semantic information, we translate key instructions from their binary format to key expressions ( self-defined symbolic expressions). At the last step in translation, we simplify the Mixed Boolean-Arithmetic (MBA) expressions to further mitigate syntactic differences caused by different compiler optimizations, compilers, source code versions, and obfuscation options. 
%We use these key expressions to generate the key-semantic graph. Each node is a key instruction, with its key expression as its node attribute.
%After graph generation, we further simplify each IR for the subsequent diffing. 
%We only translate key instructions into IR because their IRs also contain arguments and operands from non-key instructions. Therefore, the semantic information is still reserved after the translation.

\item \textbf{Graph Diffing}. It has four major steps.
First, we serialize a key-semantics graph into a sequence of key expressions by topological sort. 
Second, we tokenize each key expression to produce a list of token sequences. 
Third, we concatenate all the token sequences for all the key expressions and use the locality-sensitive hash (LSH) to hash the concatenated tokens and generate an LSH hash value for one function. 
Last, we diff two given functions by quantifying the \texttt{Jaccard} similarity between two generated LSH hash values. 
\end{itemize}

By doing so, a similarity score will be computed for the selected pair of functions. For each function in \texttt{BinA}, we do a 1-to-n compare with all the functions in \texttt{BinB} and pick a pair of functions that has the highest score as the most similar one. If the picked pair has the same function name, it means that we detect the correct pair. 

\textcolor{black}{Note that, compared to GNN-based methods, \name does not require collecting a significant amount of training data or performing time-consuming pretraining. This alleviates the challenge of obtaining abundant, high-quality assembly code for training in real-world applications.}

\subsection{Graph Generation}\label{sec:graph_gen}
%In this section, we first introduce how to extract symbolic expressions of all key instructions in a function by symbolic execution. We then translate the extracted symbolic expressions into key expressions. Last, we form a key-semantics graph based on the given function's control flow. 
%Key-semantics graph generation contains two processes. The first one is to generate key instructions' symbolic expressions. Then these symbolic expressions are input to the second process to simplify MBA expressions and generate a key-semantics graph.
%\subsubsection{Key Instruction Symbolic Expression Extraction}
%\subsubsection{Extracting Symbolic Expressions of Key instructions}

\subsubsection{Key-Instruction
Symbolic Expression
Extraction}
%\subsubsection{Extracting Key Instructions}
To efficiently extract all symbolic expressions of key instructions from a given function, we customize symbolic execution by proposing two techniques. %In the first technique, we perform a 
%a complete code-coverage and efficient
%to extract key instructions from a given function. 
%To generate the above-mentioned key-semantics graph, 
%We design a customized symbolic execution method for two purposes.
{First}, we symbolically execute a function to traverse all its instructions. {Second}, we symbolically execute a loop in a lightweight way rather than repeatedly executing the loop till the loop condition is not satisfied.
%to boost the process. 
%The output is key instruction's symbolic expression.

\mypara{Traversing All Instructions in A Function}
For a given function in a binary, we perform a complete instruction traversal %using random walk 
as shown in \autoref{alg:code coverage}. The input for the algorithm is a function's first instruction. The function is regarded as a control-flow graph where a node denotes an instruction and children of the node are subsequent instructions conforming to the function's control flow.
At the beginning of the algorithm, we also need to provide symbolic values to the function's input arguments. Particularly, we assign symbolic values $VAR_i$ ($i \in N$) to relevant registers.
The order for assigning values for the registers is based on the x86-64 calling conventions. For example, on x64 Linux, register \texttt{rdi,rsi, rdx, rcx} represents arguments 1 to 4 of the function.
%regardless of whether it is key instruction or not. 
%in the form of a graph. This graph differs from the key-semantics graph. 

%The process of complete instruction traverse are shown in \autoref{alg:code coverage}. The input is the first instruction (node) of a function. A function is treated as a graph. Each instruction corresponds to a node. Note that the graph here differs from the key-semantics graph as the graph here contains all instructions in the function regardless of whether the instruction is key instruction or not. 
The \autoref{alg:code coverage} implements a depth-first searching function, i.e., \textit{execute\_next\_node}. 
which achieves a complete instruction coverage rather than a complete code path to avoid the serious path-explosion problem.
Specifically, we first check whether a node has been executed before (Line 2). If no, we symbolically execute the node (Line 3). As an instruction in a node can have one or more operands, each of its operands will produce a symbolic expression after the symbolic execution. Thus, we maintain a record of symbolic expressions for each operand in each instruction. If the data to be referenced is resolvable (e.g., \texttt{mov esi, address} where \texttt{address} points to a string of \texttt{``Rtmin''}), we use its resolved value (\texttt{``Rtmin''}) to continue the symbolic execution. If data is unresolvable (e.g., \texttt{mov edi, cs:bio\_err} where \texttt{cs:bio\_err} is unknown), we assign unused $VAR_i$ ($i \in N$) to represent unknown values.
\textcolor{black}{IDA Pro often recognizes when an immediate value or constant represents a memory address and renames it to a variable, such as a variable name, as shown in the example above. In cases where IDA Pro cannot effectively recognize that a constant is an address—due to address hiding (e.g., \textbf{lea + add} instructions) or obfuscation—our method includes an additional step to compare whether the calculated constant falls within the address space. If it does, \name treats it as an address by default. As a result, the constant value is not used for comparison; instead, a symbolic expression, such as \textbf{VAR3}, is used for comparison.}

If the node has been executed before, we check whether the node is the start of a loop (Line 8). If yes, we process the loop in a lightweight way, i.e., $lightweight\_loop\_processing()$ shown in Line 10 and discussed in \textit{Lightweight Loop Processing} later.
%loop processing method was introduced later to process the loop. 
%This process corresponds to the function $lightweight\_loop\_processing()$ function. 
If the node has been executed before but does not form a loop, we simply return (Line 11) to avoid repeated node execution. 

\begin{algorithm}
\footnotesize
\DontPrintSemicolon
%\KwInput{$Func$, a binary function in form of a graph.}
%\SetKwFunction{FMain}{complete\_traverse}
%  \SetKwProg{Fn}{Function}{:}{}
%  \Fn{\FMain{$Function$}}{
%        $i\gets Function.start\_node$\;
%        $execute\_next\_node(i)$\;
%       % \KwRet\;
%  }
 \SetKwFunction{Function}{execute\_next\_node}
  \SetKwProg{Fn}{Function}{:}{}
  \Fn{\Function{$Node$}}{
  
  \If{Node has not been executed before}{
  $symbolic\_execution(Node)$\; \tcp{symbolically execute the instruction and update relevant records of symbolic expressions.}%\zhi{symbolic?} 
  \ForEach{$j\in Node.children$}{$execute\_next\_node(j)$\;}
%       \uIf{Node has only one child %\zhi{what if n has no child?}\Zian{solved}
%       }
%{
%    $i\gets Node.child$\;
%    $execute\_next\_node(i)$\;
%}
%\uElseIf{Node has more than one child}{
%    $i\gets random\_child(Node)$\;
%    $execute\_next\_node(i)$\;
%    \ForEach{$j\in Node.children, j\neq i$}{$execute\_next\_node(j)$\;}
%}
%\Else{\KwRet\;}
    %\KwRet\;
    }
    \ElseIf{$Node$ forms a loop}{
    
    $loop=extract\_loop(Node)$\;
    $lightweight\_loop\_processing(loop)$\;}
    %\Else{\KwRet\;}
    
  }

%$i\gets Func.start\_node$\;\tcp{$start\_node$ is the first instruction of $Func$}
        
%$execute\_next\_node(i)$\;
  
\caption{Complete Instruction Traversal}
\label{alg:code coverage}
\end{algorithm}

%\begin{figure}
%\centering
%\centerline{\includegraphics[width=50mm]{code coverage.pdf}}
%\includegraphics[width=0.7\columnwidth]{code coverage.pdf}
%\caption{Example of code coverage}
%\label{code coverage}
%\end{figure}

\begin{figure}[h]
\centering
\includegraphics[scale=0.5]{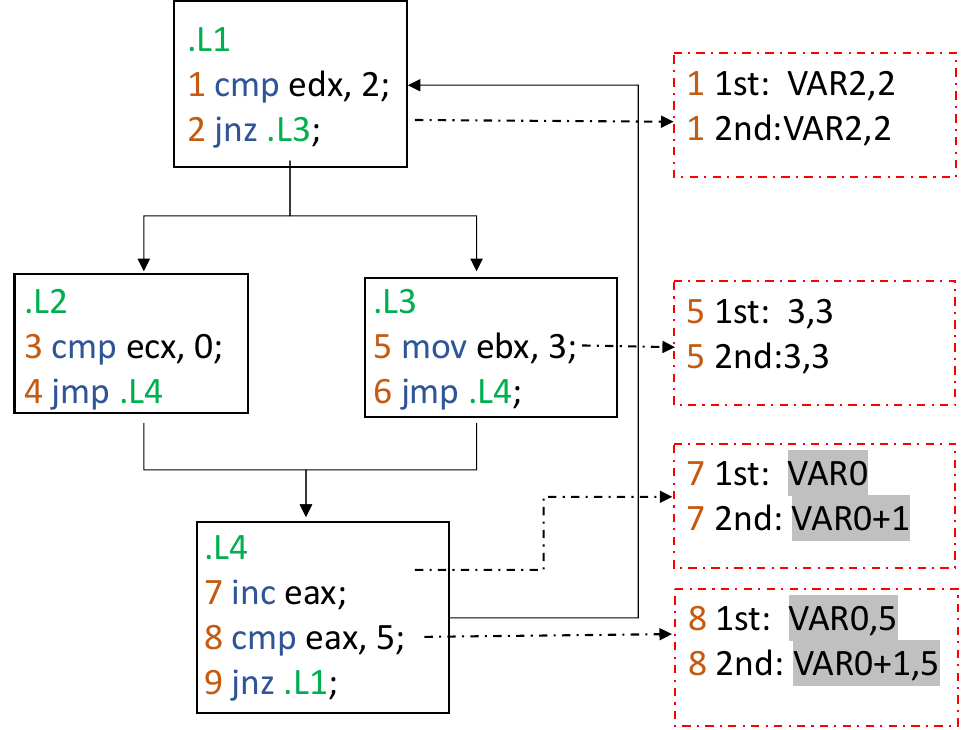}
\caption{An example of a loop.}
\label{fig:loop1}
\end{figure}

\mypara{Lightweight Loop Processing}
\label{sec:meth:loop}
In the symbolic execution above, it is inefficient to execute a loop as the loop can be repeated many times or even infinite. Instead, we propose a lightweight approach. Considering that a loop updates one or more variables each time (e.g., adding or subtracting a counter value), we call such a variable a \emph{loop counter}. When we encounter a loop, we execute the loop only two times. 
%process the loop before continuing symbolic execution. To process the loop, we 
{If there are branches within the loop, we randomly select one branch at the first time of loop execution. In the second time loop execution, we follow the same path. For example, in \autoref{fig:loop1}, there is more than one branch in the loop. In the first time of execution, we randomly select a path ${.L1-.L3-.L4}$. We execute ${.L1-.L3-.L4}$ and their symbolic expressions are denoted in \autoref{fig:loop1} after each \texttt{1st:} symbol. If there is more than one operand, the symbolic expressions of the operands are separated by a comma. For example, \texttt{1st: 3,3} at line 5 in $L3$ means that, after the symbolic execution, the first operand's symbolic expression is 3, and the second operand's symbolic expression is also 3. In the second time execution, we follow the same path. Their symbolic expressions are denoted after each \texttt{2nd:} symbol}. We then compare each operand's symbolic expression after the first and second execution to detect operands with changed symbolic expressions. We add a symbol \texttt{ITER} as the prefix to it, meaning it is a loop counter. For example, in \autoref{fig:loop1}, we identify \texttt{eax} in lines 7 and 8 as loop counter because the first time the symbolic expression of \texttt{eax} is \textit{VAR0} and the second time the symbolic expression of \texttt{eax} is \texttt{VAR0+1}. This means this variable increase by 1 in each iteration. Therefore, we change \texttt{eax}'s symbolic expression to \texttt{ITER(VAR0)}. %\Zian{For example, in block .L4 line 1, after the first execution, the symbolic value of \texttt{eax} is $VAR0+1$. After the second execution, \texttt{eax} has value $VAR0+1+1$. We can instantly detect \texttt{eax} as a loop variable and assigns it as ITER $(VAR0+1)$.}

%Then we continue the symbolic execution.
%For example, the blue blocks in Fig.\ref{loop} are loop blocks. We ignore all the other instructions within these blocks. However, in Block 4, there are two instructions. \textit{add rax, 20h} adds 0x20 to register \textit{rax} per loop. \textit{mov [rcx], rax} then copies the value in \textit{rax} to the memory pointed by \textit{rcx}. Since \textit{rax} is updated per loop, we call it loop variable. \textit{rcx} is invariant regardless of time of loop. Thus it is not a loop variable. 
%\begin{figure}
%\centering
%\centerline{\includegraphics[width=30mm]{loop.pdf}}
%\includegraphics[width=0.4\columnwidth]{loop.pdf}
%\caption{Example of loop}
%\label{loop}
%\end{figure}

%. As such, the simplified IR better represent a key code behavior.
%As the last step in key instruction translation, 
%\zhi{As a sequence of data calculating instructions with different syntax but same semantic can produce same symbolic expression after simplification,}
%we  simplify each translated IR. \zhi{how, example?} 
%\zhi{We use the APIs in \texttt{msynth}~\cite{msynth} to mathematically simplify a symbolic expression.}
%The rationale is that  
%is the memory address to write. \textit{Exp\_2} represents the value to write.

%The aforementioned two modules transform a function into an attributed IR-based graph. 
 %Thus, we now need to match or diff two attributed graphs, i.e., $G_A=(V_A,E_A)$ and $G_B=(V_B,E_B)$ where {$V_A=\{V_{A_1},...,V_{A_n}\}$ and $V_B=\{V_{B_1},...V_{B_n}\}$} are nodes and each node is a key instruction. $E_A=\{(i,j)\mid i,j\in V^2_A\}$ and $E_B=\{(i',j')\mid i',j'\in V^2_B\}$ represent the control flow among key instructions.  
%$Attr_A=\{Attr_{A_1},...,Attr_{A_n}\}$ and %$Attr_B=\{Attr_{B_1},...,Attr_{B_n}\}$ are attributes of nodes and each represents a key instruction's IR. 

\subsubsection{Key Expressions
Translation}
%\textcolor{black}{Remark: Yes, this subsection does not have any novelty, and should agree to the reviewer to embed it into the main body.}

As symbolic Expressions of Key Instructions have been extracted, they can be complicated and still have many syntactic differences due to compiling techniques such as data encoding~\cite{data_encoding} and Mixed-Boolean-Arithmetic~\cite{MBA}.
As such, we further translate them into key expressions in two steps. 
%The previous subsection generate key instruction's symbolic expressions. However, they are not the final key expression since they can be 1) overly complex from simple expressions or constant values,  or 2) syntactically different symbolic expressions due to different compiling settings. 
%To generate a key expression, 
First, we use expression-synthesizing techniques~\cite{david2020qsynth,deobfuscation1,kan2019automated} to synthesize a symbolic expression into a simplified one, e.g., $x\lor y -x\land y \Rightarrow x\bigoplus y$. This technique can effectively transform long and complex symbolic expressions into simpler and shorter expressions.
Second, we translate each key instruction to key expression according to the rules as shown in \autoref{tab:IR}.
%Second, .
%which can recover the original expression version or an expression with close semantically equivalence. 
%An example of symbolic expression synthesizing is: . Obviously 
%We use an open source python project msynth~\cite{msynth} to synthesize the symbolic expressions. Lastly we connect the key expressions according to their control flow to form the key-semantics graph.
%\mypara{Symbolic Expression Simplification}
%We further simplify MBA expressions in each translated key expression %using \texttt{msynth}~\cite{msynth}
%for better similarity detection. The MBA simplification aims at mitigating the impact caused by different compiling optimizations, different compilers, different source code versions, and obfuscation. Since in these cases we may retrieve obvious different MBA expressions for key expressions having similar semantics.  \zhi{no high-level idea about this simplification?}

\subsubsection{Generating A Key-Semantics Graph}
\begin{table*}[!ht]
\centering
\footnotesize
\resizebox{\textwidth}{20mm}{\begin{tabular}{c|c|c}
\hline
Key-Instruction Type & Key Expression & A List of Token Sequences\\
\hline
\multirow{2}{*}{calling behavior}  & \multirow{2}{*}{$RET\_FuncAddr(exp\_1,...,exp\_n)$}&$RET\_(exp\_1\_token\_1),..., RET\_(exp\_1\_token\_n), ...;$\\
&&$RET\_(exp\_n\_token\_1),..., RET\_(exp\_n\_token\_n)$\\
\hline
\multirow{2}{*}{comparing manner} & \multirow{2}{*}{$exp\_1\; cmp\; exp\_2$} & $cmp\; exp\_1\_token\_1, ..., cmp\; exp\_1\_token\_n, ...$;\\
&&$cmp\; exp\_2\_token1, ..., cmp\; exp\_2\_token\_n$\\
\hline
indirect branch & $branch\; exp$ & $branch\; exp\_token\_1, ..., branch\; exp\_token\_n$\\
\hline
\multirow{2}{*}{memory store} & \multirow{2}{*}{$[exp\_1]\; = \; exp\_2$} & $[exp\_1
\_token\_1]=, ...[exp\_1
\_token\_n]=;$\\
&&$=exp\_2\_token\_1, ...,=exp\_2\_token\_n$\\
%&&$=exp\_n\_token\_1, ...,=exp\_n\_token\_1$\\
\hline
\end{tabular}
}\caption{Key expression tokenization.}
\label{tab:token_table}
\end{table*}
%\subsubsection{Graph Generation}
%A key instruction is a node in the key-semantic graph. 

\begin{figure*}
\centering
\includegraphics[width=\textwidth]{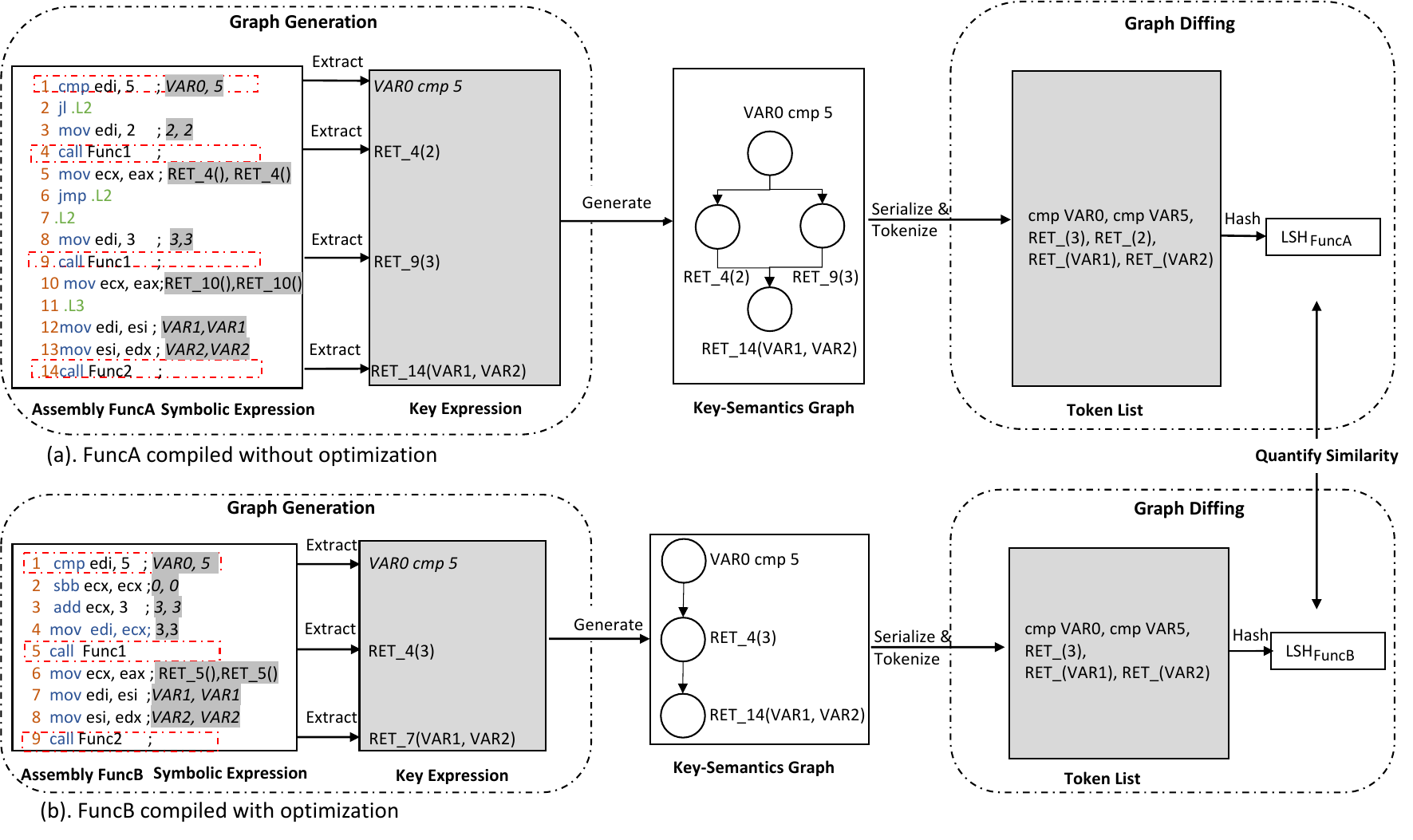}
\caption{An example of (key-semantics) graph generation and graph diffing. FuncA and FuncB are compiled from the same source code without and with optimizations.}
\label{fig:graph_generation}
\end{figure*}

Key instructions in a function have been turned into key expressions. However, until now, the key instructions are still an unsorted list, as highlighted in red rectangles in \autoref{fig:graph_generation}. Now we connect the key expressions based on a function's control flow to produce a key-semantics graph. Each node in the key-semantic graph represents a key instruction, represented as a vertex $V=\{V_{1},...,V_{n}\}$ in the key-semantics graph. The vertex attribute $Attr_i$ ($i\in \{1, ..., n\}$) is its key expression. The edges $E=\{(i,j)\mid i,j\in V^2\}$ in the graph represent the control-flow among key instructions.
%A key-semantics graph preserves key semantics (i.e., key code behaviors) in a function of a binary. It consists of key instructions as vertexes $V=\{V_{1},...,V_{n}\}$, and control-flow relationship among key instructions as edges $E=\{(i,j)\mid i,j\in V^2\}$. Each node $V_i$ has an attribute $Attr_i$ ($i\in \{1, ..., n\}$), which represents a \emph{key expression} derived from a key instruction. We note that there is no edge attribute. 

For example, \autoref{fig:graph_generation} shows how graph generation processes a pair of binary functions compiled from the same source code with and without optimization. 
%\name processes stripped binary but for better demonstration we show a non-stripped binary. %All key instructions are highlighted in light blue. Their corresponding key expressions are demonstrated in green at the end of each line.%A four-digit (on the left) in \autoref{list:key_IR_example_asm} represents a memory address of an instruction and all key instructions are highlighted in light blue in Line 23C7, 23CD, 23DF, 23E8, respectively. Their corresponding key expressions are demonstrated in green at the end of each line.%To demonstrate the difference between our Key IR graph with the control flow graph, we list both of them. %The corresponding control flow graph is demonstrated in \autoref{fig:key IR example CFG}. This is directly output from IDA pro. \autoref{fig:key IR example graph} demonstrates the corresponding Key IR  graph. 
%As an example of key-semantics graph generation, in \autoref{fig:graph_generation}, 
Specifically, it first extracts symbolic expressions of key instructions from a given binary function and then translates the extracted expressions into key expressions. Last, it connects the key expressions based on the function's control flow to generate a key-semantics graph.%Their attributes  are $[53108B]\; cmp\; 0$, $RETURN\_23CD()$, $VAR\textsubscript{0}\; cmp\; 0$, and $VAR0()$. 

\subsection{Graph Diffing}\label{sec:module3}
%\zhi{Graph Diffing Engine?}\Zian{solved}
%In this section, we introduce the third module, i.e., graph diffing.
%The aforementioned two modules transform a function into an attributed IR-based graph. For the graph, each node corresponds to a key instruction and a node attribute is a translated and simplified IR. Thus, we now need to match or diff two attributed graphs, i.e., $G_A=(V_A,E_A)$ and $G_B=(V_B,E_B)$ where {$V_A=\{V_{A_1},...,V_{A_n}\}$ and $V_B=\{V_{B_1},...V_{B_n}\}$} are nodes and each node is a key instruction. $E_A=\{(i,j)\mid i,j\in V^2_A\}$ and $E_B=\{(i',j')\mid i',j'\in V^2_B\}$ represent the control flow among key instructions.  
%$Attr_A=\{Attr_{A_1},...,Attr_{A_n}\}$ and %$Attr_B=\{Attr_{B_1},...,Attr_{B_n}\}$ are attributes of nodes and each represents a key instruction's IR. 
%To this end, We compute the similarity between $G_A$ and $G_B$ based on both geometric and attribute similarity. 

%existing solutions (e.g., Siglidis et al.~\cite{grakel} and Morris et al.~\cite{morris2019weisfeiler}) diff the geometric relations between the graphs. However, these solutions only diff a limited number of node neighbours by geometric relations due to computational complexity, and they require that the attributes of a graph should be vectorized.
To quantify the similarity between the two given graphs, existing approaches vectorize the attribute of each node. Thus, this requires the attributes must be short symbols. 
However, the attributes of our key-semantic graph can be long expressions, making existing approaches inapplicable. %they cannot be trivially converted to vectors and we need a new solution. %\Zian{(because the attribute can be )}, 
%we cannot leverage the existing solutions, 
%\Zian{which assumes the attribute to be short and easy to be embedded}. 

Inspired by ~\cite{codesensor,guanjunlin_tdsc} where source codes are transformed into a sequence of instructions to traverse an Abstract Syntax Tree in a linear order, we address the problem in three steps. 1) We serialize a key-semantics graph into a sequence of nodes by \emph{topological sort}. 2) We tokenize the key expressions (i.e., each node {attribute} in the serialized graph). 3) We concatenate the tokenized key expressions and apply \emph{locality-sensitive hash} (LSH) to produce an LSH value for similarity quantification.

\subsubsection{Key-Expression
Sequences serialization}\label{graph_serialization}
The technique of {topological sort} can sort all the nodes $V=\{V_1,...,V_n\}$ of a directed graph $G=(V,E)$ in a linear order and thus $G$ satisfies the following property: every directed edge $\{(V_u,V_v)\mid V_u,V_v\in V^2\}$ is forward, i.e., $V_u$ comes before $V_v$. More specifically, topological sort can keep the structural and geometrical relations among the nodes. Similar graphs can result in similar topological sequences. %Therefore, we can use \emph{topological sort} to serialize a graph. 

However, it is possible that a loop in the function contains multiple key instructions. Therefore, graph generation step can produce loops in key-semantics graph. And topological sort works only for a directed graph that has no loop, which cannot be directly applied to a key-semantics graph that can contain loops. %inherited from a function (e.g., a loop in a function has key instructions). 
To address this issue, we first make a key-semantics graph loop-free by removing the last edge in the flow of a loop and then adding a symbol of $WHILE$ into the starting node of the loop, meaning it is the beginning of the loop. As shown in \autoref{fig:loop1}, $.L4$ to $.L1$ is the last edge of the flow and is removed. The nodes in the loop body are retained.

%we use a symbol called $WHILE$. Based on the control flo, we remove the last edge along the control flow in the loop to make the graph loop-free. We then .{ An example of the last edge along the control flow within the loop is shown as  in . This is because we need to retain the complete loop body and its control flow. }
%\zhi{I'm lost here. We need a more detailed description}.\Zian{How should we elaborate this? A picture?} %Fig.\autoref{topological sort} shows an example of this process. The red symbols denote changes.

%\begin{figure}
%\centering
%\centerline{\includegraphics[width=40mm]{pictures/Key_IR_example_graph.pdf}}
%\includegraphics[width=0.5\columnwidth]{pictures/topological_sort.pdf}
%\caption{Delete loops in the Key IR graph}
%\label{topological sort}
%\end{figure}

\subsubsection{Key-Expressions
Tokenization}
%We now generate a sequence of instructions for each IR graph. 
%To diff two serialized graphs, we further transform each serialized graph into a sequence of key expression tokens and use {locality sensitive hash (LSH)} for token-based sequence diffing. 
%Specifically,
For each node {attribute} in the serialized graph, we split its key expression into operands and operators, constituting a sequence of tokens,
%Thus, we omit the plus symbol in tokens. 
e.g., ${X+3-Y+7*Z}$ is spit and tokenized into a sequence of ${X, 3, -, 7, *, Z}$.
If a key expression has brackets or parentheses, we retain the brackets or parentheses for each token when splitting it, e.g., parts of two key expressions are $[X+[Y+Z-3]*2]$ and $(X+(Y*6+(Z-3+K)))$. They are tokenized as: $[X], [[Y]], [[Z]], [[-]], [[3]], [*], [2]$ and $(X), ((Y)), ((*)), ((6)), (((Z))), (((-))),$\newline $ (((3))) (((K)))$. 
%By doing so, a generated token sequence retains more semantic information. 
We note that the plus symbol (${+}$) in a key expression is omitted when it is split, as we observe that a tokenized plus symbol dominates a token sequence, which makes two different token sequences have a high similarity score. 
To associate a token with its corresponding key-expression type, we prefix a token with a symbol as shown in ~\autoref{tab:token_table}.
%and processed 
%For an IR, we As ,  .have the following processing rules:
%\zhi{is this also a tool? no reference?}\Zian{an algorithm, other paper referencing this LSH have no reference}. 
%However, directly feeding the sequence of tokens is infeasible because of two reasons. First, 
%\zhi{is this based on your empirical observation}\Zian{solved} %\textbf{2)} For \textbf{type 1} Key instruction, calling subfunction, since the binary is stripeed, the function name is replaced with an address. However same function names does not imply same addresses. The address is independent of the function name. \textbf{3)} Still for the same type of Key instruction, the function parameters can be confused with non-parameter symbolic expressions. \textbf{4)} For \textbf{type 2} Key instruction, comparison, comparing operand can be confused with non-comparing symbolic expressions. \textbf{5)} For \textbf{type 3} Key instructions, memory modify, offset within the bracket can be confused with non-offset symbolic expressions. Likewise, \textbf{6)}, operands within a parentheses can be confused with a symbolic expression not in the same parentheses.
%Second, the way how tokens are split also affects the similarity score. 

%To this end, we need an effective method to split instructions into tokens.  Even though there are many possible ways to split the instruction, %we propose a heuristic method here: 

\begin{itemize}[noitemsep, topsep=2pt, partopsep=0pt,leftmargin=0.4cm]
%Two examples are shown in \autoref{list:split_example_bracket} and \autoref{list:split-example_parentheses}.%\autoref{common sym1} and \autoref{common sym2}. 
\item For a calling behavior, %$exp_i$ ($i\in \{1, ..., n\}$) in its key expression is split and constitutes a token sequence. 
each token is prefixed with $RET\_()$, denoting that it comes from a calling instruction, e.g., a key expression is $RET\_(3,VAR0+4)$ becomes $RET\_(3), RET\_(VAR0), RET(4)$.
%is within a \texttt{RETURN\_()} symbol. Then we split each parameter as a common symbolic expression, each token with a \texttt{RETURN\_()} symbol outside of it, as shown in \autoref{tab:token_table}. This denotes the type of token belongs to a calling convention instruction. 

\item For a comparing manner, %$exp_i$ ($i\in \{1, 2\}$) as its comparing operands generates a token sequence, and 
each token is prefixed with $cmp$, e.g., a key expression $4\;cmp\;[VAR1+18]$ turn into $cmp\;4$, $cmp\;[VAR1]$, $cmp \;[18]$.
%into two comparing operands, each with a \texttt{cmp} as the prefix. Then we split the two tokens as two common symbolic expressions, each with a \texttt{cmp} symbol as the prefix, as \autoref{tab:token_table}. 

\item For an indirect branch, %we split $exp$ into a token sequence, and 
each token is prefixed with $branch$, e.g., $branch\;[[VAR2+10]+16]$ turns into $branch\;[[VAR2]]$, $branch\;[[10]]$, $branch\;[16]$.

\item For a memory store, %$[exp\_1]$ is split into a token sequence and each is with ${=}$ as the suffix. 
tokens on the left of equal sign ends with ${=}$, %For $exp\_2$, its token is with ${=}$ as the prefix, 
tokens on the right of the equal sign prefixed with ${=}$,
e.g., $[VAR2+18]=(VAR1+10)*3$ becomes $[VAR2]=,[18]=,=(VAR1),=(10),=*,=3$.
\end{itemize}
\eat{
\begin{figure}[h]
\centering
\includegraphics[width=0.4\textwidth]{pictures/graph_diffing.pdf}
\caption{key expression tokenization}
\label{fig:graph_diffing}
\end{figure}
}

\subsubsection{LSH Values Hashing and Quantifying Similarity}
%As exact nearest neighbor is not scalable to handle high dimensional data, he LSH approach has been used to solve the approximated Nearest Neighbor problem. One can 
The LSH approach is effective for nearest neighbor search in high-dimensional spaces as LSH can represent high-dimensional data in a low-dimension format while preserving the relative distance between data. 
Particularly, this approach hashes data objects into buckets so that similar objects will be hashed into the same bucket with a high probability.
%\zhi{Thus LSH algorithm ensures sub-quadratic space complexity and a provably sub-linear query time.}
As such, we utilize LSH to generate an LSH value for a serialized and tokenized key-semantics graph. Specifically, we first concatenate all the token sequences from all the key expressions in a serialized key-semantics graph. We then generate an LSH value for the concatenated token sequence to represent one function. With two LSH values from two given functions, we use the \emph{Jaccard similarity} to quantify the similarity between the two functions, as the Jaccard similarity shown below is commonly used to measure similarity or distance for two given datasets. 
$$J(A,B)=\dfrac{\left|A\cap B\right|}{\left|A\cup B\right|} ,$$
where $A$ and $B$ are two datasets, and the number of the common elements in them is divided by their total number.

\autoref{fig:graph_generation} shows an example of how graph diffing operates. It first serializes two given key-semantics graphs and then tokenizes the key expressions, which are concatenated to generate a token list. Last, it computes LSH values for token lists from each binary function and quantifies their similarity.

\textcolor{black}{In our implementation, we use Python's \textbf{networkx} library's topological sort API, along with \textbf{datasketch's Minhash}, to serialize the key-semantics graph into a token list and hash it.}
%as shown below.
%it can be used to quantify the similarity between two 
%of two sets (i.e., two sets of tokenized key expressions in our case).

%\autoref{fig:graph_diffing} shows two examples of key expression tokenization from example in  \autoref{fig:graph_generation}. Then the tokenized key expression of all the nodes in one key-semantics graph is concatenated together to produce a LSH hash value. Later one can use the LSH hash values to diff them. 

\mypara{Source Lines of Code}
Our implementation for the first module contains 12,527 \texttt{C++} SLOC and it serves as a plugin to \texttt{IDA Pro}. The second module has 2,913 python SLOC. We use APIs provided by \texttt{msynth}\footnote{https://github.com/mrphrazer/msynth} as a python package to synthesize symbolic expressions of key instructions.

%Our implementation for the third module has 2,913 python SLOC. 

%\begin{listing}
%Original expression: 

%$[X+[Y+Z-3]*2]$
%\newline

%Processed expression:

%$[X], [[Y]], [[Z]], [[-]], [[3]], [*], [2]$
%\caption{Example for processing brackets}
%\label{list:split_example_bracket}
%\end{listing}

%\begin{listing}
%Original expression: 

%$(X+(Y*6+(Z-3+K)))$
%\newline

%Processed expression:

%$(X), ((Y)), ((*)), ((6)), (((Z))), (((-))), (((3))) (((K)))$
%\caption{Example for processing parentheses}
%\label{list:split-example_parentheses}
%\end{listing}

%$$RETURN\_FunctionAddress(exp\_1,...,exp\_n)$$
%\begin{equation}
%\label{type1_token}
%\Downarrow
%\end{equation}
%$$RETURN\_(exp\_1,..., RETURN\_(exp\_n$$

%$$exp\_1\quad cmp \quad exp\_2$$
%\begin{equation}
%\label{type2_token}
%\Downarrow
%\end{equation}
%$$cmp(exp\_1, cmp(exp\_n$$

%$$[exp\_1]\quad = \quad exp\_2$$
%\begin{equation}
%\label{type3_token}
%\Downarrow
%\end{equation}
%$$[exp\_1]=, =exp\_n$$

\section{Evaluation}
\eat{\begin{figure}[h!]
\centering
%\centerline{\includegraphics[width=100mm]{pictures/key_IR_CFG.pdf}}
%\includegraphics[width=\columnwidth]{pictures/gcc_ncd.pdf}
\includegraphics[scale=0.20]{pictures/gcc_ncd.pdf}
\caption{Each open-source program is compiled by five selected pairs of GCC optimization levels, and an NCD score for each pair of binaries is presented.}
\label{fig:gcc_ncd}
\end{figure}
}
%\section{Evaluation}\label{sec:eva}

\label{sec:experiment}

We evaluated the effectiveness of \name answering the following research questions (\textbf{RQs}):
\begin{itemize}
\item \textbf{RQ1.} What is the accuracy of the generated key expressions?
\item \textbf{RQ2.} Can \name detect similarity across different compiling optimizations?
\item \textbf{RQ3.} Can \name detect similarity across different compilers?
\item \textbf{RQ4.} Can \name detect similarity across different obfuscations?
\item \textbf{RQ5.} Can \name be applied to cross-version similarity detection?
\item \textbf{RQ6.} How does \name perform in vulnerability detection?
\item \textcolor{black}{\textbf{RQ7.} Can the proposed key instructions effectively capture the program semantics?}
\item \textcolor{black}{\textbf{RQ8.} Does assigning different weights to the various key instructions affect the accuracy?}
\end{itemize}

\subsection{Experiment Setup}

\noindent
\textbf{Dataset}
\label{experiment:data}
%\zhi{the reason why we choose the following programs is lost}
We conducted experiments on nine popular open-source projects, i.e.,
\tool{openssl-3.3.0}, \tool{libtomcrypt-1.18.2}, \tool{coreutils-8.32}, \tool{ImageMagick-7.1.010}, \newline \tool{libgmp-6.2.1}, \tool{curl-7.80}, \tool{sqlite3-3.37.0}, \tool{zlib-1.2.11} and \tool{Puttygen-0.74}.
From these libraries, 
    we selected 13 programs (listed in~\autoref{tab:gcc}) that are widely used by vendors and the other researchers~\cite{asm2vec, IMF-SIM, qbindiff}. The source code of these programs consists of 52 project packages. The compiled binary code include 2106 binaries, with more than 416,000 assembly functions due to different compilers, compiling options, source code versions, and obfuscations.
%as shown in ~\autoref{tab:gcc}.
%as shown in ~\autoref{tab:gcc_clang_stat}. 
%These programs are widely used by vendors and are also used by previous similarity detection works~\cite{asm2vec, IMF-SIM, qbindiff}. 
%Note that Coreutils originally contains approximately 150 binaries. We modify the build configuration to generate Coreutils binaries as a single binary. 
%In particular, Coreutils is compiled into about 150 binaries by default and we modify its build configuration to merge them into a single binary. 
%\Zian{We make more than 150 binaries in Coreutils into a single binary for simplicity.}
%libMagickCore and libMagickWand are from the Imagemagick project, while plink, pscp, psftp, and puttygen are from the Puttygen project. For libtomcrypt, we can compile it into a single ELF executable or a shared library. 

\noindent
\textbf{Environment}
We set up a machine with an Intel NUC8i5BEH (Intel processor i5-8259U with 16\,GB memory).
Since the experiment might involve deep learning, 
    we also use an accelerator cluster of HPC systems with 456 NVidia Tesla P100, 114 Dual Xeon 14-core E5-2690 v4 compute nodes and 256\,GB memory.
    
%Since some baseline tools need deep learning, we prepare an accelerator cluster of HPC systems with 456 NVidia Tesla P100, 114 Dual Xeon 14-core E5-2690 v4 compute nodes and 256\,GB memory. The other tools are launched on an Intel NUC8i5BEH (Intel processor i5-8259U with 16\,GB memory).

\noindent
\textbf{Experiment}
We compare \name with five state-of-the-art tools of binary similarity detection (i.e., \tool{JTrans}~\cite{jtrans}, \tool{UniASM}~\cite{uniasm}, \tool{Asm2Vec}~\cite{asm2vec}, \tool{Gemini}~\cite{Gemini}, and \tool{Palmtree}~\cite{Palmtree}). %BinDiff is a widely used product from the industry and the other 4 tools are representative academic works.
%We downloaded each existing tool from their public source, i.e.,  BinDiff {Version 5} (BinDiff {V5})~\cite{bindiff}\footnote{https://zynamics.com/software.html}, functionsimsearch~\cite{functionsimsearch}\footnote{https://github.com/googleprojectzero/functionsimsearch}, Asm2Vec~\cite{asm2vec}\footnote{https://github.com/McGill-DMaS/Kam1n0-Community}, Gemini~\cite{Gemini}\footnote{https://github.com/xiaojunxu/dnn-binary-code-similarity}, Palmtree~\cite{Palmtree}\footnote{https://github.com/palmtreemodel/PalmTree}. 
We trained \tool{JTrans}~\cite{jtrans}, \tool{UniASM}~\cite{uniasm}, \tool{Gemini} and \tool{Palmtree} on the HPC, and validate all the tools on Intel NUC. %and other tools on an Intel NUC. 
%For \tool{BinDiff}, we utilize the function level features to achieve similarity comparison.
\tool{JTrans}, \tool{functionsimsearch}, \tool{Gemini}, and \tool{Palmtree} are machine learning based approaches.
Hence, we leverage a total of %12 binaries 
7000 assembly functions from four programs (i.e., \tool{busybox}, \tool{coreutils}, \tool{libgmp}, and \tool{libMagickCore}) compiled with different optimization levels for model training. 
Aligned with existing works~\cite{asm2vec,discovre,genius}, we consider functions with at least five basic blocks as functions with less than 5 blocks are less likely to contain bugs, and thus are of less interest for similarity detection. 
%consider functions that have at least five basic blocks because functions with fewer blocks have fewer possibilities to contain vulnerability thus of less interest for similarity detection.

%BinDiff by default uses different features to detect binary similarity in the function level, such as function-name match, which is however often unavailable for real-world stripped binaries. Thus, we refrain from using its function-name match and retain its other features. 
%As functionsimsearch, Gemini and Palmtree are based on supervised learning, we provide four programs (i.e., busybox, coreutils, libgmp, and libMagickCore) as their training dataset and use the aforementioned 13 programs as their test dataset. {We note that coreutils, libgmp, and libMagickCore are included in both the training and test dataset, they can be used to evaluate the detection performance of the  learning-based tools on similar data. The other 10 programs in the test dataset can used to test their performance on unseen data, which is realistic in real-world binary similarity detection.} 
%Aligned with existing works~\cite{asm2vec,discovre,genius}, we consider functions that have at least five basic blocks because functions with less blocks has much less possibilities to contain vulnerability thus of less interest for similarity detection.

\noindent
\textbf{Similarity Measurement}
Given a pair of binary codes (i.e., \texttt{BinA} and \texttt{BinB}) compiled by different optimization levels, we compute the similarity score as below.
First, 
    we select a function from \texttt{BinA} and execute a binary similarity detection tool to acquire a similarity score between the selected function and each function in \texttt{BinB}.
The pair of functions with the largest score is regarded as the most similar function.
Then we enable the debugging information of the function based on the function name to check whether functions in \texttt{BinB} and \texttt{BinA} are the same.
%Specifically, for each function in \texttt{BinA}, each tool acquires its similarity score with every function in \texttt{BinB} and selects a pair of functions with the largest score as the most similar function. To verify whether each selected pair is from the same function, we enable the debugging information of function names when compiling a program, i.e., when the selected pair passes the function-name check, a correct paired function is found in \texttt{BinB} for \texttt{BinA}.  
To quantify the similarity, we use the same metrics as the previous works~\cite{asm2vec,unleashing}.
That is precision at position 1 (precision@1), which captures the ratio of binary functions from \texttt{BinA} that correctly find the function with the same name in \texttt{BinB} at position 1. Precision@1 is equal to Recall at Position 1 (Recall@1) in this case.

%We assess \name in terms of detecting ground-truth similar functions and its real-world applications. The following questions are answered: %As the same source code can generate different binary code when different compiling optimization levels, compilers, source-code versions or obfuscation option are used~\cite{bindiff,functionsimsearch,asm2vec,Gemini,Palmtree}, we compile the collected programs into binaries using a list of configurations of cross-GCC-compiling-optimization-level in~\autoref{sec:cross_optimization_sec}, cross-compiler in~\autoref{sec:cross_compiler_sec}, cross-program-version in~\autoref{sec:cross_version_sec} and different obfuscation options in~\autoref{sec:cross_obfuscation}, respectively. Detecting binary similarity in these configurations is beneficial to real-world scenarios such as 1-day vulnerability detection.

%\Zian{We aim at testing whether the binary can still be detected as similar under those different configurations. As these are close to realistic scenarios such as program linage, 1-day vulnerability detection, etc.} 
 %Last, we measure the running overhead of \name in~\autoref{sec:performance}, which is relevant to the binary size. 
Please note that in the following sections, we use the aforementioned program versions for evaluation. For ~\autoref{sec:cross_version_sec}, we use the aforementioned program version as the baseline and compared them with other versions in terms of their binary similarity, e.g.,  curl-3.3.0 is against curl-7.43, curl-7.65 and curl-7.72. 

\subsection{RQ1: Correctness of Key Expressions}

\begin{table}[h]
\centering
\footnotesize
\scalebox{0.9}{\begin{tabular}{c|c|c|c}
 \hline
  Project  & Functions (\#)  &Project  & Functions (\#) \\ 
  \hline
Coreutils-O0 & 25 & Curl-O1 & 30\\ 
Coreutils-O3 & 20 & libgmp-Os & 31\\
Coreutils-Os & 27 & libmagickcore-o1 &24\\
libtomcrypt-O1 & 23 & libmagickcore-o2 &20\\

 \hline

\end{tabular}}
\caption{Statistics of the analyzed functions.We randomly selected 200 functions in total from various projects with different optimization levels.}
\label{tab:correctness}
\end{table}

To justify the correctness of the translated key expressions, we randomly selected 200 functions from the dataset described in \autoref{experiment:data} 
and asked three experienced programmers to label them manually. 
%For each function, we read each binary instruction line and check the correctness of their symbolic expressions in order to check the correctness of the key 
The 200 functions are shown in \autoref{tab:correctness}. It demonstrated that \name can correctly translate 85\% of instructions. 
By manually inspecting the incorrect cases, 
    we observed that \name cannot recognize (currently not supported) some x64 mnemonics' variants (e.g., \texttt{movzx}) that are less frequently used in binary codes. 
As \name is built on top of IDA pro to resolve string variable names into contents of the string,
    IDA pro might mistakenly resolve the strings.
For example, %in a legitimate case, \name resolves string variable \texttt{address} from instruction \texttt{mov esi, address} into \texttt{``Rtmin''} where \texttt{address} points to. However, 
IDA pro may consider constant value as a memory address and resolve the content in that memory address.

\begin{table*}
\footnotesize
%\begin{tabular}{p{1.4cm}p{0.7cm}p{0.7cm}p{0.7cm}p{0.7cm}p{0.7cm}p{0.7cm}}
\resizebox{\textwidth}{25mm}{\begin{tabular}{c| c | c | c | c | c | c }
 \hline
\multirow{2}{*}{Program} & JTrans & UniASM & Asm2Vec & Gemini & Palmtree& \name\\%\multicolumn{6}{c}{($O1$ vs. $O0$, $O3$ vs. $O0$, $Os$ vs. $O0$)} \\
  &O1-O0 O3-O0 Os-O0&O1-O0 O3-O0 Os-O0&O1-O0 O3-O0 Os-O0&O1-O0 O3-O0 Os-O0&O1-O0 O3-O0 Os-O0&O1-O0 O3-O0 Os-O0  \\\hline
 openssl &(0.715,0.658,0.631) & (0.712,0.642,0.650) & (0.798, 0.724, 0.733) & (0.320, 0.272, 0.246) & (0.368, 0.319, 0.328) & (\textbf{0.818, 0.779, 0.759})\\ 
 %\hline
 libtomcrypt & (0.694,0.697,0.685) & (0.756,0.730,0.698) & (0.673, 0.604, 0.700) & (0.075, 0.034, 0.059) & (0.124, 0.063, 0.117) &(\textbf{0.800, 0.766, 0.716}) \\
 %\hline
 coreutils & (\textbf{0.571},0.432,0.416)  &(0.523,0.468,\textbf{0.642})  & (\textbf{0.570}, 0.416, 0.525) & (0.169, 0.144, 0.171) & (0.285, 0.179, 0.254) &(0.553, 0.479, 0.625)\\
 %\hline
 libMagickCore & (0.729,0.674,0.658)  &(0.712,0.703,0.641)  &(0.648, \textbf{0.729}, \textbf{0.721}) & (0.166, 0.126, 0.127) & (0.249, 0.197, 0.189) & (\textbf{0.740}, {0.715}, 0.664)\\
 %\hline
 libMagickWand & (0.851,0.837,0.728)  & (0.902,0.861,0.795)  & (0.459, 0.734, 0.804) & (0.076, 0.075, 0.056) & (0.160, 0.121, 0.151) &(\textbf{0.930, 0.875, 0.812})\\ 
 %\hline
 libgmp & (0.739,0.752,0.682)  & (0.752,0.781,0.718)  & (0.705, \textbf{0.823}, 0.733) & (0.485, 0.505, 0.409) & (0.660, 0.738, 0.587) &(\textbf{0.783}, 0.808, \textbf{0.746})\\ 
 %\hline
 curl & (0.802,0.702,0.756)  &(\textbf{0.881,0.757},0.809)  & (0.753, 0.699, 0.703) & (0.362, 0.372, 0.333) & (0.529, 0.394, 0.505) &(0.825, 0.731, \textbf{0.791})\\ 
 %\hline
 sqlite3 & (\textbf{0.751},0.508,0.675)  & (0.709,\textbf{0.531},0.685)  & (0.687, 0.433, 0.640) & (0.132, 0.061, 0.128) &(0.229, 0.111, 0.206) &(0.738, 0.522, \textbf{0.695})\\ 
 %\hline
 libz & (0.729,0.627,0.785)  & (0.757,0.659,0.819) & (0.767, \textbf{0.750}, 0.836) & (0.297, 0.224, 0.206) & (0.525, 0.259, 0.402) &{(\textbf{0.778}, 0.681, \textbf{0.849}})\\ 
 %\hline
 plink & (0.689,0.507,0.653)  & (0.677,0.512,0.672)  & (0.690, 0.473, 0.673) & (0.263, 0.229, 0.213) & (0.372, 0.250, 0.323) &(\textbf{0.748, 0.562, 0.800})\\ 
 %\hline
 pscp & (0.673,0.516,0.689) & (0.680,0.523,0.674)  & (0.679, 0.406, 0.634) & (0.244, 0.206, 0.244) &(0.379, 0.241, 0.302) &(\textbf{0.754, 0.551, 0.811})\\ 
 %\hline
 psftp & (0.637,0.509,0.657)  & (0.689,0.527,0.671)  & (0.690, 0.439, 0.640) & (0.248, 0.210, 0.240) & (0.361, 0.253, 0.334) &(\textbf{0.767, 0.559, 0.787})\\ 
 %\hline
 puttygen & (0.615,0.457,0.623) & (0.632,0.461,0.663)  & (0.648, 0.371, 0.624) &(0.261, 0.184, 0.263) & (0.379, 0.247, 0.374)
 &(\textbf{0.673, 0.513, 0.770})\\ 
 \hline
 Avg. &(0.707,0.605,0.664)  & (0.721,0.627,0.702) & (0.674, 0.585, 0.690)  & (0.238, 0.203, 0.207) & (0.355, 0.259, 0.313) & (\textbf{0.762, 0.657, 0.756})\\ 
 \hline

\end{tabular}
}\caption{Libraries compiled by GCC5.4.0 with different optimization levels. funcsimsrch is short for functionsimsearch.}
\label{tab:gcc}
\end{table*}

\begin{table*}[h]
\footnotesize
\resizebox{\textwidth}{25mm}{\begin{tabular}{c| c | c | c | c | c | c }
%{p{1.4cm}p{0.7cm}p{0.7cm}p{0.7cm}p{0.7cm}p{0.7cm}p{0.7cm}}
 \hline
\multirow{2}{*}{Program} & JTrans & UniASM & Asm2Vec &Gemini & Palmtree & \name\\%\multicolumn{6}{c}{(CLANG $O0$ vs. GCC $O0$, CLANG $O3$ vs. GCC $O3$)} \\
  &O0-O0 O3-O3&O0-O0 O3-O3&O0-O0 O3-O3&O0-O0 O3-O3&O0-O0 O3-O3&O0-O0 O3-O3  \\ 
 \hline
 openssl & (0.723,0.809)  & (0.741,0.838)  & (\textbf{0.841}, 0.852) & (0.358, 0.232) & (0.181, 0.265) & (0.770, \textbf{0.877})\\ 
 %\hline
 libtomcrypt & (0.603,0.727)  & (0.632,0.752)  &(0.625, \textbf{0.844}) & (0.125, 0.124) & (0.127, 0.090) & (\textbf{0.758}, 0.770)\\ 
 %\hline
 coreutils & (0.708,0.817)  & (0.720,\textbf{0.838})  & (0.695, 0.797) & (0.222, 0.181) & (0.216, 0.280) &(\textbf{0.752}, 0.828) \\ 
 %\hline
 libMagickCore & (0.605,0.749)  & (0.627,0.771)  & (0.607, \textbf{0.862}) & (0.216, 0.175) & (0.204, 0.210) & (\textbf{0.757}, 0.812)\\ 
 %\hline
 libMagickWand & (0.642,0.785)  & (0.682,0.837)  & (0.324, 0.795) & (0.08, 0.133) & (0.054, 0.070) & (\textbf{0.872}, \textbf{0.912})\\ 
 %\hline
 libgmp & (0.472,0.582)  & (\textbf{0.495},0.612)  & (0.376, 0.639) & (0.135, 0.249) & (0.150, 0.277) & (0.487, \textbf{0.661})\\ 
 %\hline
 curl & (0.805,0.775)  & (0.812,0.763)  & (0.845, 0.725) & (0.404, 0.356) & (0.284, 0.322) & (\textbf{0.938}, \textbf{0.900})\\ 
 %\hline
 sqlite3 & (0.729,0.628)  & (0.753,0.602)  & (0.795, 0.617) & (0.191, 0.109) & (0.119, 0.156) & (\textbf{0.890}, \textbf{0.697}) \\ 
 %\hline
 libz & (0.723,0.675) & (0.781,0.709)  & (0.863, 0.728) & (0.555, 0.267) & (0.336, 0.395) & (\textbf{0.961}, \textbf{0.827})\\ 
 %\hline
 plink & (0.708,0.687)  & (0.721,0.706)  & (0.764, 0.748) & (0.234, 0.230) & (0.165, 0.297) & (\textbf{0.870}, \textbf{0.820})\\ 
 %\hline
 pscp & (0.715,0.693)  & (0.731,0.712)  & (0.752, 0.743) & (0.242, 0.199) & (0.155, 0.241) & (\textbf{0.879}, \textbf{0.819})\\ 
 %\hline
 psftp & (0.637,0.723)  & (0.659,0.745)  & (0.644, 0.755) & (0.261, 0.207) & (0.174, 0.262) & (\textbf{0.878}, \textbf{0.820})\\ 
 %\hline
 puttygen & (0.714,0.619)  & (0.730,0.689)  & (0.762, 0.636) & (0.292, 0.182) & (0.242, 0.246) & (\textbf{0.870}, \textbf{0.767})\\ 
 \hline
 Avg. & (0.675,0.713)  & (0.698,0.736)  & (0.694, 0.749) & (0.255, 0.203) & (0.185, 0.239) & (\textbf{0.823}, \textbf{0.808})\\ 
 \hline
\end{tabular}
}\caption{Binary comparison result when using different compilers. funcsimsrch is short for functionsimsearch.}
\label{tab:clang_same_optimization}
\end{table*}

\subsection{RQ 2: Cross-GCC-Compiling-Optimization-Level Similarity Detection}\label{sec:cross_optimization_sec}
%\Zian{To answer RQ1, we we quantify the similarity of multiple binaries compiled from the same source code with different compiling optimization levels. Even though there are many compilers we can choose from, we select GCC because it is most widely used.} 
GCC is one of the most widely used compilers in the real world and thus we choose it for cross-compiling-optimization-level evaluation. 
%\zhi{use one or two sentences to describe the popularity of GCC}\Zian{no need, other papers do not mention gcc popularity}
We conduct the experiment by setting five optimization levels, $O0$, $O1$, $O2$, $O3$, and $Os$, and apply \name and the other five state-of-the-art tools to detect the similarity of the binary codes generated from the same source function, but compiled by using different optimization levels.
In addition, 
    we leverage the metrics of Normalized Compression Distance (NCD) score~\cite{unleashing,AlshahwanBCDM20,10.1145/3097983.3098111,Borbely2015OnNC,10.1145/3128572.3140446} to quantify the syntactic similarity of the binary code pair.
A higher NVD score represents the binary code pair looks more dissimilar. 

The similarity detection results are listed in~\autoref{tab:gcc}.
Due to the page limitation, we only list the detection results of the most dissimilar pair (i.e., $O3$ vs $O0$) and the two other dissimilar pairs (i.e., $O1$ vs $O0$ and $Os$ vs $O0$).
The other results and the corresponding discussions are all listed on our website \textcolor{black}{\footnote{https://anonymous.4open.science/r/SemDiff4BinaryDetection-F12C/README.md}}. \textcolor{black}{Note that -O2 is not included in the comparison because it largely overlaps with -O1 and -O3, and therefore is not expected to produce qualitatively different behavior than those already presented. In GCC, both -O3 and -Os have significant overlap with -O2.~\autoref{tab:various_optimization_levels} provides a more detailed comparison of the different optimization levels. Each optimization level consists of a set of specific optimization flags.}
The similarity detection results illustrate that \name averagely achieves a detection precision at 73\%, 
    but the other detection tools can only achieve a detection prediction from 18\% to 74.9\%.%at 65\% on average.
Although \tool{JTrans, UniASM, Asm2Vec} slightly performs better than \name in some cases, 
    \name could maintain a high detection performance even though the pair of binary codes look the most dissimilar.

\begin{table}[h!]
\begin{tabular}{c|ccc}
Compared levels & Common flags & Extra Flags       & Removed/Different \\
\hline
O0 vs O1        & $\sim$0      & +30$\sim$35 in O1 & None              \\
O1 vs O2        & $\sim$30     & +25 in O2         & None              \\
O2 vs O3        & $\sim$60     & +10$\sim$15 in O3 & None              \\
O2 vs Os        & $\sim$45-50  & +5$\sim$10        & $\sim$10-15      
\end{tabular}
\caption{Comparison of different optimization levels}
\label{tab:various_optimization_levels}
\end{table}

\subsection{RQ 3: Cross-Compiler Similarity Detection}\label{sec:cross_compiler_sec}
Similarly, 
    we conduct the experiment by using two different compilers (i.e., \tool{GCC 5.4.0} and \tool{CLANG 3.8.0}), but the same optimization level to assess the similarity detection performance of \name.
By checking the NVD score of the binary code pairs, 
    we found that \tool{GCC} $O0$ vs \tool{CLANG} $O0$ and \tool{GCC} $O1$ vs \tool{Clang} $O1$ are the two most dissimilar binary code pairs. 
Therefore, 
    we listed the similarity detection results in the paper (shown in~\autoref{tab:clang_same_optimization}).
The rest results are listed on our website as well. 

On average, 
    \name could achieve a detection precision at 81\% when using the same optimization level, but different compilers. 
Also in some cases, \tool{JTrans}, \tool{UniASM}, and \tool{Asm2Vec} performs slightly better than \name.

For binaries compiled from the same source code using different optimization levels in \autoref{sec:cross_optimization_sec} or different compilers in \autoref{sec:cross_compiler_sec}, their function numbers vary mainly due to the \texttt{inline} optimization, which inserts functions being called into the callee function. Also, 
they are likely to differ in almost all the binary functions as some instructions inside a function have syntactic differences but the same semantics. These differences will result in different function attributes such as the statistics of basic blocks, instructions, and mnemonics. Thus methods that rely on syntactic information (all except \name) are less accurate. However, most key semantics of a function is still preserved in this case, making \name more effective than other tools. 
%as shown in \autoref{tab:gcc_stat} and \autoref{tab:clang_stat}. 
%Within the changed function, some of the instructions are substitute with same semantic meanings. 
%This can hugely influence the accuracy of the methods based on basic information such as basic block numbers, instruction numbers, or statistics of each mnemonic. 
%\zhi{Some optimization flags also inline functions and modify control flows.} 
%This significantly change the structure of the function. But the functions still have the same semantic meaning, i.e., the same behaviour and impact during execution. This implies the invarience of the Key IR.

%evaluate CLANG $O0$ vs. GCC $O0$ (~\autoref{tab:clang o0 gcc o0}), CLANG O1 versus GCC O0 ( ~\autoref{tab:clang o1 gcc o0}), CLANG O3 versus GCC O0 ( ~\autoref{tab:clang o3 gcc o0}), CLANG O3 versus GCC O1 ( ~\autoref{tab:clang o3 gcc o1}), CLANG O3 versus GCC o3 ( ~\autoref{tab:clang o3 gcc o3}). We report the average result. Generally, \name outperforms all the baselines. 

\begin{table}[h]
\centering
\footnotesize
\begin{tabular}{p{0.5cm}cp{0.5cm}p{0.5cm}p{0.5cm}p{0.5cm}p{0.5cm}p{0.5cm}}
 \hline
  Program & Version & \rotatebox[origin=c]{270}{JTrans} & \rotatebox[origin=c]{270}{UniASM} & \rotatebox[origin=c]{270}{Asm2Vec} & \rotatebox[origin=c]{270}{Gemini} & \rotatebox[origin=c]{270}{Palmtree} & \rotatebox[origin=c]{270}{\name}  \\ 
  \hline
\multirow{3}{*}{openssl} & \multicolumn{1}{r}{1.0.2o vs. 3.0.0}& 0.522 &0.463 &0.428 &0.284 &0.382 &\textbf{0.534} \\ 
   & \multicolumn{1}{r}{1.1.1i vs. 3.0.0} &0.753 &0.652 &0.839 &0.576 &0.649 &\textbf{0.918} \\ 
  & \multicolumn{1}{r}{1.1.1 vs. 3.0.0}& 0.667 &0.676 &0.813 & 0.568 &0.648&\textbf{0.923} \\ 
 \hline
\multirow{3}{*}{coreutils} & \multicolumn{1}{r}{8.27 vs. 8.32} & 0.859 &0.825 &\textbf{0.895} &0.551 &0.626 &0.877
  \\ 
  &  \multicolumn{1}{r}{8.29 vs. 8.32} & 0.856 &0.761 &\textbf{0.899}  &0.576 &0.671 &0.878
  \\ 
  & \multicolumn{1}{r}{8.31 vs. 8.32}&  \textbf{0.932} &0.897 &0.916 &0.591 &0.673 &0.900
  \\ \hline
\multirow{3}{*}{curl} & \multicolumn{1}{r}{7.43 vs. 7.80} &0.697 &0.598 &0.800 &0.438 &0.603 &\textbf{0.908}\\
  & \multicolumn{1}{r}{7.65 vs. 7.80} & 0.890 &0.827 &0.795 & 0.509 &0.741&\textbf{0.955}\\
  & \multicolumn{1}{r}{7.72 vs. 7.80} & {0.904} &0.764 &0.822  & 0.588 &0.743&\textbf{0.946}\\
  \hline
\multirow{3}{*}{libz} & \multicolumn{1}{r}{1.2.3.4 vs.1.2.11} & 0.632 &0.697 &0.700 &0.520 &0.560&\textbf{0.800}\\
  & \multicolumn{1}{r}{1.2.3.9 vs. 1.2.11} & 0.637  & 0.612  &0.786 &0.599 &0.598 &\textbf{0.833}\\ 
  & \multicolumn{1}{r}{1.2.8 vs. 1.2.11} & 0.751 &0.732 &0.839 & 0.747 &0.768&\textbf{0.903} \\
 \eat{& \multicolumn{1}{r}{1.18.0rc3 vs. 1.18.2} &0.783 &0.280 &0.739 & 0.409 &0.469&\textbf{0.856} \\ 
 %\hline
  libtcrpt &  \multicolumn{1}{r}{1.18.0 vs. 1.18.2} & 0.776 &0.276 &0.736 & 0.434 &0.446&\textbf{0.863}\\ 
 %\hline
  & \multicolumn{1}{r}{1.18.1 vs. 1.18.2}& 0.785 &0.275 &0.750 &0.418 &0.467 &\textbf{0.869} \\ 
 \hline
 Avg.&  &0.781  &0.277  &0.742  &0.420 &0.460 &\textbf{0.863}\\ 
 \hline
 }
 
 \hline
 Avg.& & 0.758	&0.708	&0.779&0.548	&0.658	&\textbf{0.870}\\
 \hline

\end{tabular}
\caption{The similarity scores are computed by 6 tools for open-source programs of different versions compiled by GCC 5.4.0. (funcsimsrch is short for functionsimsearch.)}
\label{tab:cross_version}
\end{table}

\subsection{RQ 4: Different Obfuscation Options Similarity Detection}\label{sec:cross_obfuscation}
We also quantify the similarity detection performance by using different obfuscation options, 
    that is two programs that are compiled by \tool{CLANG 3.8.0} with $O0$ and \tool{OLLVM}~\cite{ollvm} with three different obfuscation options (i.e., SUB, BCF, and FLA). %\zhi{Based on our observation?}, 
\tool{OLLVM} with SUB substitutes specific instructions within a basic block and does not change a function's control-flow graph significantly. With BCF, it adds additional basic blocks into a function in a fixed pattern. With FLA, it flattens the control flow by simply splitting original basic blocks into smaller ones or adding extra basic blocks into an original function. %\Zian{The data versions are the same as \autoref{sec:cross_optimization_sec} and \autoref{sec:cross_compiler_sec}.}

%We choose the same binaries from \autoref{sec:cross_optimization_sec} for compilation, 
For each program, we generate three pairs and each pair consists of a CLANG-compiled binary and an OLLVM-compiled binary with one obfuscation option. We then fed each pair into \name and two representative machine-learning-based tools (i.e., Gemini and Palmtree) for similarity quantification. The results are shown in \autoref{tab:obfuscate_sub_bcf}. Clearly, \name outperforms the other two tools for all obfuscation options by large margin. 
\eat{
\begin{table}[htp]
\centering
\footnotesize
\begin{tabular}{c | c c | c c | c c}
\hline
\multirow{2}{*}{Program} & \multicolumn{2}{c}{OLLVM-SUB} & \multicolumn{2}{c|}{OLLVM-BCF} & \multicolumn{2}{c}{OLLVM-FLA}\\
 %\hline
  & Palmtree & \multicolumn{1}{c}{\name} & Palmtree & \multicolumn{1}{c}{\name} & Palmtree & \multicolumn{1}{c}{\name}\\  
 \hline
 curl & 0.248 &\textbf{0.967}
  & 0.021 &\textbf{0.922}
 &  0.021 &\textbf{0.75}
\\ 
 %\hline
 sqlite3 &  0.113 &\textbf{0.814} & 0.007 &\textbf{0.941} & 0.001 &\textbf{0.639}
\\
 %\hline
 libz  & 0.355 &\textbf{0.941} & 0.073 &\textbf{0.963} & 0.027 &\textbf{0.817}
\\
 %\hline
 plink  &  0.161 &\textbf{0.940} & 0.014 &\textbf{0.644} &  0.002 &\textbf{0.694}
\\
 %\hline
 pscp  & 0.170 &\textbf{0.945} & 0.019 &\textbf{0.653} & 0.003 &\textbf{0.689}
\\ 
 %\hline
 psftp  & 0.145 &\textbf{0.936} & 0.014 &\textbf{0.656} &  0.003 &\textbf{0.688}
\\ 
 %\hline
 puttygen & 0.195 &\textbf{0.930} & 0.032 &\textbf{0.671} & 0.011 &\textbf{0.718}
\\ 
 \hline
 Avg. & 0.198 &\textbf{0.925} & 0.026 &\textbf{0.779} & 0.010 &\textbf{0.714}\\ 
 \hline
\end{tabular}

\caption{The similarity scores are computed by Palmtree and \name for open-source programs compiled by CLANG 3.8.0 with $O0$ and OLLVM with different obfuscation options.}
\label{tab:obfuscate_sub_bcf}
\end{table}
}

\begin{table}[h]
\centering
\footnotesize
%\setlength\tabcolsep{3.0pt}
%\begin{tabular}{c c c c c c c c  c c c}
\resizebox{\textwidth}{34mm}{\begin{tabular}{c |ccc|ccc|ccc}
\hline
\multirow{2}{*}{Program} & \multicolumn{3}{c}{OLLVM-SUB} & \multicolumn{3}{c}{OLLVM-BCF} & \multicolumn{3}{c}{OLLVM-FLA}\\
 %\hline
  &\rotatebox[origin=c]{270}{Gemini}& \rotatebox[origin=c]{270}{Palmtree} & \rotatebox[origin=c]{270}{\name} & \rotatebox[origin=c]{270}{Gemini} & \rotatebox[origin=c]{270}{Palmtree} & \rotatebox[origin=c]{270}{\name} &\rotatebox[origin=c]{270}{Gemini} & \rotatebox[origin=c]{270}{Palmtree} & \rotatebox[origin=c]{270}{\name}\\  
 \hline
  openssl &0.627  &0.713  &\textbf{0.973} &0.034 &0.056 &\textbf{0.840} &0.004 &0.004 &\textbf{0.444}
  \\
  libtomcrypt & 0.309 &0.520 &\textbf{0.973} &0.021 &0.026 &\textbf{0.849} &0.003 &0.004 &\textbf{0.721}
 \\
  coreutils & 0.410 & 0.589 &\textbf{0.835} &0.033 &0.037 &\textbf{0.839} & 0.001 &0.001 &\textbf{0.582}
  \\
  libMagickCore &0.403 &0.481 &\textbf{0.934} &0.011 &0.018 &\textbf{0.885} &0.001 &0.001 &\textbf{0.623}
 \\
  libMagickWand &0.192 &0.129 &\textbf{0.997} &0.027 &0.032 &\textbf{0.962} &0.001 &0.006 &\textbf{0.733}
  \\
  libgmp &0.310 &0.649 &\textbf{0.769} &0.073 &0.085 &\textbf{0.715} &0.061 &0.066 &\textbf{0.604}
\\ 
 curl &0.348 &0.248 &\textbf{0.967}
  &0.055 &0.021 &\textbf{0.922}
 & 0.014 &0.021 &\textbf{0.75}
\\ 
 %\hline
 sqlite3 & 0.113 & 0.113 &\textbf{0.814} &0.011 &0.007 &\textbf{0.941} &0.001 &0.001 &\textbf{0.639}
\\
 %\hline
 libz  &0.400 &0.355 &\textbf{0.941} & 0.073 &0.073 &\textbf{0.963} &0.027 &0.027 &\textbf{0.817}
\\
 %\hline
 plink  & 0.178 &0.161 &\textbf{0.940} &0.019 &0.014 &\textbf{0.644} & 0.001 &0.002 &\textbf{0.694}
\\
 %\hline
 pscp  &0.188 &0.170 &\textbf{0.945} &0.018 &0.019 &\textbf{0.653} &0.001 &0.003 &\textbf{0.689}
\\ 
 %\hline
 psftp  &0.186 &0.145 &\textbf{0.936} &0.019 &0.014 &\textbf{0.656} & 0.001 &0.003 &\textbf{0.688}
\\ 
 %\hline
 puttygen &0.226 &0.195 &\textbf{0.930} &0.026 &0.032 &\textbf{0.671} & 0.006 &0.011 &\textbf{0.718}
\\ 
 %\hline
 Avg. &0.234 &0.198 &\textbf{0.925} &0.032 &0.026 &\textbf{0.779} &0.007 &0.010 &\textbf{0.714}\\ 
 \hline
\end{tabular}
}\caption{The similarity scores of Gemini, Palmtree and \name for programs compiled by CLANG 3.8.0 with $O0$ and OLLVM with different obfuscation options.}
\label{tab:obfuscate_sub_bcf}
\end{table}

%\Zian{The main reason is possibly because that obfuscation impact heavily on the syntactical and structural information but does not alter the binary semantics.} 
 
%For the experiments in \autoref{sec:cross_optimization_sec} and \autoref{sec:cross_compiler_sec}, we suspect \name can abstract higher level of semantic information into key expression from the plain binary instruction and the LSH hashing can effectively compare the two key-semantic graphs with both topological and semantic information. Therefore \name achieves the best results. In comparison, all other tools directly take the binary code as the input, which can contain more noise with obscure semantic information thus decreases the results.

To understand the root cause of the failure cases of \name, we manually analyzed the results and found that in experiments of \autoref{sec:cross_optimization_sec} and \autoref{sec:cross_compiler_sec}, when \name failed to rank the ground truth similar function at the first place, in approximately 50\% of the cases, \name still rank the similar function before 10th place. We consider this still can assist humans to find similar functions efficiently. In the other 50\% cases, \name failed to rank similar functions at front positions mainly due to three reasons: 1) Lack of support for some less frequent mnemonics such as \texttt{cvtss2sd}. This can negatively impact semantic information extraction thus decreasing precision.  2) Some calls are optimized into other instructions. For example, \texttt{call strlen} is replaced with \texttt{repne scasb}, which has the same impact and behavior as \texttt{call strlen}. Even using symbolic execution, their symbolic values still differ enormously. 3) Sometimes, the unfolded loop and the folded loop can be difficult to match. Because their symbolic expressions can differ. And the loop number does not match (i.e., only one unfolded loop but multiple unfolded loops).

For the experiment in \autoref{sec:cross_obfuscation}, we speculate that although the obfuscation options obfuscate a binary in terms of its syntactic structures, they retain its key semantics, which can be retrieved by \name.  
%most key semantics from binaries, which are retained even after obfuscation. 
For the three evaluated tools, their generated scores under the SUB option achieve the highest compared to the other options. This is probably because the SUB option 
does not change the control flow.
Of the three options, scores in the FLA option are the lowest, as it introduces more syntactical and control-flow changes by flattening the control flow. The failure cases are caused by newly added key instructions by obfuscation, despite the three reasons mentioned in the last paragraph.
\subsection{RQ 5: Cross-Program-Version Similarity Detection}\label{sec:cross_version_sec}
%In this experiment, we evaluate binaries compiled from different versions. 
%This experiment evaluates the precision of \name and baseline tools on binaries of different versions. 
In real-world applications, binary code similarity can be utilized to find similar versions of library binaries or executables because vulnerabilities tend to inherit across a range of versions. Therefore, in this section, for two given binaries from the same program with different versions, we quantify their similarities. 
%We select the same binaries as \autoref{sec:cross_optimization_sec} and \autoref{sec:cross_compiler_sec}. 
For each program, four versions spanning from months to years are compiled using GCC 5.4.0. For each program, one version is selected as the baseline version (we note that this version is used in previous experiments of \autoref{sec:cross_optimization_sec} and \autoref{sec:cross_compiler_sec}.) and its similarity with each of the other three versions is computed using precision@1, generating {39} pairs of binaries in total. Part of the results is displayed in ~\autoref{tab:cross_version} due to the page limit. Please refer to our GitHub repository for complete results.

Overall, \name outperforms all the other tools. 
{Particularly, \name achieves the best detection performance in 10 programs and ranks second in the remaining 3 programs, i.e., coreutils, libgmp,  and sqlite3. For both coreutils and sqlite3, \name's averaged score is only 0.01 lower than that of Bindiff. For libgmp, \name's score is only 0.03 lower than that of Asm2vec.} A possible reason why \name performs less well in the 3 programs is: as the version difference in the 3 programs is smaller than that of the 10 programs, it indicates that the versions in these programs have more similarities in syntactic structures, which are easier to be captured by tools that rely on syntactic and structural features. When the version difference becomes larger in other programs, \name performs the best. The reasons for the failure cases in this experiment are also mainly due to lack of support for rare mnemonics, replacing calls to equivalent instructions, and difficulty in precisely matching loops.

\subsection{RQ 6: Vulnerability Detection}
An important application of binary code similarity detection is to find similar vulnerable functions. We randomly selected 40 Common Vulnerabilities and Exposures (CVEs) functions and detect their similar vulnerable functions. For each vulnerable function, we randomly select a vulnerable version of it as the base function. We also prepared another randomly selected vulnerable function version, compiled with random compiling settings (i.e., either O0, O1, O2, O3, Os). They are the target functions that the tools should detect as similar to the base function. We mix the target functions with all other functions from the binary and check the probability of the tool successfully ranking the target vulnerable function at the first place among all functions (i.e., top-1 score). The result is shown in \autoref{tab:vul}. \textcolor{black}{Gemini's top-1 score is 52.5\% (21 out of 40), UniASM's top-1 score is 42.5\% (17 out of 40), Asm2Vec's top-1 score is 75.0\% (30 out of 40), while \name is 80.0\% (32 out of 40)}. 

We manually analyzed the CVEs where \name fails to identify (rank at the first place). We found that out of 8 failure cases, in 6 cases (75\%) \name ranked the vulnerable function before 10th place. This still indicates the effectiveness of using \name to find vulnerabilities. For the other 2 failure cases, one is due to IDA pro failing to identify the indirect jump addresses thus making \name unable to produce a complete key-semantics graph. Another case was due to lack of support for less frequent mnemonics, which negatively impact the semantic information extraction and thus decreased the precision.

\subsection{\textcolor{black}{RQ 7: Key instruction semantic code coverage}}
%Theoretically, the proposed four key instructions capture the majority of the semantics. 

\textcolor{black}{To assess how much of the program semantics is captured by our proposed key instructions, we also conduct a data-flow analysis to track the propagation destination of each non-key instruction. We calculate the percentage of operand values from non-key instructions that ultimately reach a key instruction in each function. We collect the compiled binaries from all the projects in the dataset, each with different optimization levels. For each function used in similarity comparison, we perform a data-flow analysis. The results are shown in \autoref{tab:data-flow-code_coverage}.}

\begin{table}[h]
\centering
\begin{tabular}{c|cc}
   & GCC    & CLANG  \\
   \hline
o0 & 94.5\% & 94.8\% \\
o1 & 95.1\% & 95.3\% \\
o3 & 94.7\% & 95.2\% \\
os & 95.0\% & 95.2\%
\end{tabular}
\caption{Data-flow analysis of code coverage by \name}
\label{tab:data-flow-code_coverage}
\end{table}

\textcolor{black}{With different optimization levels, \name retains a similar percentage of program semantics, approximately 95\%. To investigate which types of instructions are not propagated to the key instructions, we manually analyzed the results. We found that some non-key instructions, such as prologue instructions (e.g., \texttt{push eax; push ebx}), do not propagate their data into the key instructions. Other missed instructions are rarely used, such as \texttt{MULX, MOVS}, and \texttt{LSL}. Since \name is currently a proof of concept, it would be straightforward to incorporate these rarely used instructions in future versions, potentially improving code coverage.}

\subsection{\textcolor{black}{RQ 8: Weighted LSH accuracy}}
\textcolor{black}{When comparing similarity, \name uses Locality-Sensitive Hashing (LSH). By default, LSH treats each key instruction type with equal weight. However, to explore the impact of assigning different weights to different key instructions, we use weighted LSH for similarity comparison. Specifically, we assign different weights to the symbolic expression tokens of various key instruction types, while leaving the weights of the other key instructions at the default value of 1.0. We randomly select over 10,000 functions from the dataset, which includes functions compiled with different settings, compilers, and versions. }

\begin{table}[h]
\centering
\begin{tabular}{c|ccccc}
        & call & compare & indirect & store & accuracy \\
        \hline
weights & 1    & 1       & 1        & 1     & 78.15\%   \\
weights & 2    & 1       & 1        & 1     & 77.79\%   \\
weights & 1    & 2       & 1        & 1     & 78.13\%   \\
weights & 1    & 1       & 2        & 1     & 77.82\%   \\
weights & 1    & 1       & 1        & 2     & 78.13\%  
\end{tabular}
\caption{Weighted LSH similarity accuracy for different key instruction types.}
\label{tab:weighted_lsh}
\end{table}

\textcolor{black}{~\autoref{tab:weighted_lsh} shows the overall accuracy of weighted LSH on the selected data. The first row corresponds to the default LSH method, without considering weights. The results indicate that, when different weights are assigned, the accuracy remains approximately the same. This could be due to the use of symbolic execution, which mitigates the impact of low-level address differences.}

\textcolor{black}{When more weight is assigned to calling instructions or indirect branches, the accuracy decreases slightly. This may be because these instructions carry less semantic information compared to others (e.g., the majority of calling instructions contain only zero or one parameter.}

\begin{table}[h!]
\centering
\footnotesize
\begin{tabular}{ c c c c c c c c c c}
\hline
CVE &\rotatebox[origin=c]{270}{Gemini}&\rotatebox[origin=c]{270}{UniASM}&\rotatebox[origin=c]{270}{Asm2Vec}&\rotatebox[origin=c]{270}{\name} &CVE &\rotatebox[origin=c]{270}{Gemini}&\rotatebox[origin=c]{270}{UniASM}&\rotatebox[origin=c]{270}{Asm2Vec}&\rotatebox[origin=c]{270}{\name} \\
\cline{1-1}\cline{6-6}
Overflow/out of bound&&&&&Certification/Authentication&&&&\\
\cline{1-1}\cline{6-6}
CVE-2024-9143&\Checkmark&\XSolidBrush&\Checkmark&\Checkmark&CVE-2025-4575&\XSolidBrush&\XSolidBrush&\Checkmark&\Checkmark\\
CVE-2024-5535&\XSolidBrush&\XSolidBrush&\Checkmark&\Checkmark&CVE-2023-28321&\Checkmark&\XSolidBrush&\Checkmark&\Checkmark\\
CVE-2025-9086&\Checkmark&\XSolidBrush&\XSolidBrush&\Checkmark&CVE-2020-8286&\XSolidBrush&\XSolidBrush&\Checkmark&\XSolidBrush\\
CVE-2017-1000100&\XSolidBrush&\Checkmark&\Checkmark&\Checkmark&CVE-2023-27538&\XSolidBrush&\XSolidBrush&\Checkmark&\Checkmark\\
CVE-2017-1000101&\XSolidBrush&\XSolidBrush&\Checkmark&\XSolidBrush&CVE-2023-27536&\Checkmark&\XSolidBrush&\XSolidBrush&\Checkmark\\
CVE-2017-1000254&\Checkmark&\XSolidBrush&\Checkmark&\Checkmark&CVE-2023-27535&\XSolidBrush&\Checkmark&\Checkmark&\Checkmark\\
CVE-2019-5436&\Checkmark&\Checkmark&\XSolidBrush&\Checkmark&CVE-2025-49812&\Checkmark&\XSolidBrush&\Checkmark&\Checkmark\\
CVE-2017-7407&\Checkmark&\XSolidBrush&\Checkmark&\Checkmark&CVE-2017-2629&\XSolidBrush&\Checkmark&\XSolidBrush&\Checkmark\\
\cline{6-6}
CVE-2017-8817&\XSolidBrush&\XSolidBrush&\Checkmark&\XSolidBrush&Loop-related&&&&\\
\cline{6-6}
CVE-2017-1000257&\Checkmark&\Checkmark&\Checkmark&\Checkmark&CVE-2020-8285&\Checkmark&\Checkmark&\Checkmark&\Checkmark\\
 
CVE-2017-9502&\Checkmark&\XSolidBrush&\Checkmark&\Checkmark&CVE-2025-5399&\Checkmark&\Checkmark&\XSolidBrush&\Checkmark\\
\cline{1-1}  
free-related&&&&&CVE-2024-4603&\Checkmark&\Checkmark&\Checkmark&\Checkmark\\
\cline{1-1} \cline{6-6}
CVE-2024-4741&\XSolidBrush&\XSolidBrush&\Checkmark&\Checkmark&Misc&&&&\\
\cline{6-6}
CVE-2016-8618&\XSolidBrush&\Checkmark&\XSolidBrush&\XSolidBrush&CVE-2024-12797&\Checkmark&\XSolidBrush&\Checkmark&\Checkmark\\
\cline{1-1}
SSRF&&&&&CVE-2024-13176&\XSolidBrush&\XSolidBrush&\Checkmark&\Checkmark\\
\cline{1-1}
CVE-2024-43394&\Checkmark&\Checkmark&\Checkmark&\Checkmark&CVE-2024-6119&\Checkmark&\XSolidBrush&\Checkmark&\Checkmark\\
CVE-2024-10524&\XSolidBrush&\XSolidBrush&\XSolidBrush&\Checkmark&CVE-2024-2511&\Checkmark&\Checkmark&\Checkmark&\Checkmark\\
\cline{1-1}
Information leakage&&&&&CVE-2025-10148&\XSolidBrush&\Checkmark&\XSolidBrush&\Checkmark\\
\cline{1-1}
CVE-2023-46218&\XSolidBrush&\XSolidBrush&\Checkmark&\Checkmark&CVE-2023-28320&\XSolidBrush&\XSolidBrush&\Checkmark&\Checkmark\\
CVE-2023-28322&\XSolidBrush&\XSolidBrush&\XSolidBrush&\Checkmark&CVE-2016-8617&\Checkmark&\Checkmark&\Checkmark&\XSolidBrush\\
CVE-2020-8169&\XSolidBrush&\Checkmark&\Checkmark&\XSolidBrush&CVE-2016-8615&\XSolidBrush&\XSolidBrush&\XSolidBrush&\Checkmark\\
CVE-2021-22876&\Checkmark&\Checkmark&\Checkmark&\XSolidBrush&CVE-2016-8616&\Checkmark&\Checkmark&\Checkmark&\XSolidBrush\\
&&&&&CVE-2023-38546&\Checkmark&\Checkmark&\Checkmark&\Checkmark\\

\hline

\hline
\end{tabular}
\caption{Vulnerability function search results. \Checkmark represents the tool ranked vulnerable target function at the first place in all functions. \XSolidBrush is vice versa. %$^1$ and $^2$ are two functions from the same CVE ranked alphabetically.
%G, U, A, S represent Gemini, Uniasm, Asm2Vec, \name, respectively.
}
\label{tab:vul}
\end{table}

\section{Discussion}
\label{sec:discussion}

\mypara{Key Expression Accuracy}
As analyzed in \autoref{sec:experiment}, lack of support for less frequent mnemonics can decrease precision. Therefore, more accuracy can increase with more complete mnemonics supports. Due to the time limit and the complexity of all the mnemonics, our current version only supports the most frequently used mnemonics. Apart from complete mnemonics support, adding hard-coded optimization knowledge into \name can also increase the accuracy. As we analyzed before, \texttt{call strlen} and \texttt{repne scasb} has different symbolic expressions. Identifying them as the same can only be achieved through adding that optimization knowledge into \name. Since GCC and Clang are open-sourced, a promising direction is to parse those source code to learn those knowledge automatically.

\mypara{Graph Diffing}
In our current method, we use LSH algorithm to translate each graph into a hash value. Therefore, we equally consider the importance of each token in the key expression. Also, for each kind of key instruction, we also consider them to have equal importance. However, some tokens and key instructions should have more importance than others. For example, the matching of long instant values should indicate more similarity than matching some frequent operational symbols such as $*$. Matching two calling type key instructions with four arguments should weigh more than one argument. The weight of each token and key instruction type can be learned by machine learning if one prepares adequate training data.

\mypara{Inherited Limitations}
In \autoref{sec:discussion}, there were cases when the static binary code analysis platform IDA pro that we rely on failed to analyze indirect jump targets. Unfolded and folded loops matching is also not well resolved. One possible solution is to unfold the loop for a fixed time. Another solution is to conclude the unfolded loops into one loop and compare them with the folded loop. As to key expression simplification, \name utilized msynth~\cite{msynth}, which potentially has time efficiency and accuracy issues.

\section{Related Work}
\subsection{Program-analysis Based Methods}
%For syntactic and structural similarity, common strategies directly process the binary code by using different methods including hashing, embedding, and alignment. 
\textcolor{black}{SMIT~\cite{SMIT}, BINCLONE~\cite{BINCLONE}, and SPAIN~\cite{SPAIN} use hashing techniques to output various instructions sequences into a fixed length of the hash value and compare their similarity. %Equivalent output value implies syntax similarity. 
IDEA~\cite{IDEA}, MBC~\cite{MBC}, Expose~\cite{Expose} generate an embedding from sequences. Exediff~\cite{exediff}, Tracy~\cite{tracy}, and Binsequence~\cite{binsequence} align two sequences and decide their similarity. %Other common methods focusing more on structural similarity include optimization solutions, k-subgraph matching, path similarity, and graph embedding. 
SMIT~\cite{SMIT}, Binslayer~\cite{binslayer}, Cesare et al.~\cite{cxz2014} transform the problem into finding the mapping between two CFGs with minimum cost. Beagle~\cite{BEAGLE}, Cesare et al.~\cite{cxz2014}, rendezous~\cite{rendezvous}, and FOSSILE~\cite{fossil} divide the graph into $k$ subgraphs and match subgraphs similarity. %so that each one has $k$ connected nodes. Matched number of those subgraphs indicates the extent of the similarity. 
CoP~\cite{COP}, SIGMA\cite{SIGMA}, Binsequence~\cite{binsequence} determine similarity based on paths. Beagle~\cite{BEAGLE}, FOSSIL~\cite{fossil}, and SIGMA~\cite{SIGMA} classified instruction based on their arithmetic, %semantic in terms of arithmetic,
logic, or data transfer operations. Binhash~\cite{binhash}, MULTI-HM~\cite{multimh}, Bingo~\cite{bingo}, SPAIN~\cite{SPAIN}, Karg{\'e}n~\cite{ks2017}, IMF-SIM~\cite{IMF-SIM} check whether output is the same as the input. Binhunt~\cite{gao2008binhunt}, Binhash~\cite{binhash}, Expose~\cite{Expose}, CoP\cite{COP}, MULTI-MH~\cite{multimh}, ESH\cite{esh} symbolically execute the binaries and compare similarity by constraint solver. %use symbolic formulas to represent the binary code. With the symbolic formula, Binhunt~\cite{gao2008binhunt} uses Theorem Prover to check whether two different symbolic formula has similar output. Binhash~\cite{binhash}, Binjuice~\cite{binjuice}, GITZ~\cite{GITZ} use symbolic hashes as an alternative to the theorem prover. 
XMATCH~\cite{xmatch}, TEDEM~\cite{tedem} determine the edit distance of the tree/graph of the symbolic formula.} %The most representative works of this category and their limitations are introduced in \autoref{sec:intro}. 
%However, this genre of work is problematic in basic-level comparison or sequence aligning due to the difference caused by compilers and optimizations. %code-coverage problems and relies on basic-block level matching.
\textcolor{black}{Static function-level methods compare blocks or instructions.
Each block or instruction is treated as the smallest similarity unit.
However, compiler optimizations can split blocks or replace instructions.
As a result, block structures and instructions change completely.
Thus, block- or instruction-level feature, or graphical feature comparison is not ideal for binary similarity \cite{unleashing}.}

%However, this category of approaches usually fails in real-world scenarios where binaries are optimized \cite{unleashing}. 
%For example, \texttt{mov \%eax, \$0} and \texttt{xor \%eax, \%eax} have the same semantic meaning and can be replaced by each other when compiling optimization is enabled.

\subsection{Machine-learning Based Methods}
\textcolor{black}{Genius~\cite{genius}, Vulseeker\cite{VULSEEKER}, Gemini~\cite{Gemini}, Yu et al. \cite{ordermatters}, Cochard et al. \cite{graph-cross}, \textcolor{black}{TIKNIB\cite{interpretable}, Asteria-Pro~\cite{asteria}, Jason et al. \cite{jason}} extract features from graphs into feature vectors and determine the vector similarity. QBinDiff~\cite{qbindiff}extracts binary code features and uses graph edit distance and network alignment methods to measure similarity. $\alpha$Diff~\cite{aDiff}, InnerEye~\cite{innereye}, Asm2Vec~\cite{asm2vec}, Kam1n0~\cite{Kam1n0}, \textcolor{black}{J-Trans~\cite{jtrans}, Palmtree~\cite{Palmtree}, UniASM~\cite{uniasm},} and Safe~\cite{safe} automatically learn the embedding for each instruction and use them to produce the basic-level or function-level embedding.} %However, this genre of work can be affected heavily by the training data and optimization levels.
\textcolor{black}{As mentioned in section \ref{sec:intro}, 
machine-learning methods work in theory but rely on good training data.
Binaries vary widely across compilers and optimization levels [17].
This makes dataset building hard and accuracy unstable.
Unsupervised models learn from context but fail under heavy optimization.
Supervised models preserve syntax, not full semantics.}

%\subsection{Semantic similarity}
%For semantic similarity, general techniques are instruction embedding, input-output pairs, symbolic execution, theorem prover, and semantic hashes. 

\section{Conclusion}
%\begin{thebibliography}{00}
%\section{Conclusion}\label{sec:conclusion}
This paper proposed \name, which is a novel semantic method for binary similarity detection. \name has two modules: graph generation and graph diffing. In graph generation, we proposed complete instructions traversal of a given function and a lightweight loop processing to generate key instruction symbolic expressions. We further translate symbolic expressions to key expressions and form a key-semantics graph. We utilized LSH to diff two key-semantics graphs. %We 
%defined four types of key instructions
%non-key instructions, and Key-instruction graph 
%to capture critical code behaviors.
%that is, 1) calling behavior with operands from previous instructions as its parameters; 2) comparing manner with operands as its comparing objectives, 3) indirect branch with an operand as its target address, and 4) memory store with operands as its memory address and storing value. In graph diffing module, we serialize the graph using topological sort, tokenize and concatenate key expressions to token lists, and diff serialized graphs using LSH. 

In our evaluation, 
%we compiled 13 open-source programs into binaries. We evaluated cross-compiling-optimization-level similarity comparison, cross-compiler similarity comparison, cross-version similarity comparison, and the impact of binary size and function number against precision. 
We compared \name with 5 state-of-the-art baseline tools 
%(i.e., Bindiff~\cite{bindiff}, functionsimsearch~\cite{functionsimsearch}, Asm2Vec~\cite{asm2vec}, Gemini~\cite{Gemini}, and Palmtree~\cite{Palmtree}) 
in binary similarity detection, results of which showed that \name outperformed all the baseline tools on average.
While our current version of \name works for x86-based binaries
\name can be extended to support similarity detection cross architectures (e.g., x86 and ARM) in our future work.
Particularly, we will extend the graph-generation module to extract and transform key instructions from a target architecture into key expressions.
%with respect to cross-compiling-optimization-level, cross-compiler, cross-program-version, and obfuscation.
%
%Experimental results 

%\Zian{Even though we evaluated on binaries compiled from C language and GCC/Clang compiler, \name provides a general methodology to detect binary code similarity.} 
%While we evaluated on binaries compiled from C language and GCC/Clang compiler, \name provides a general methodology to detect binary code similarity

%a possible way is to transform  into Intermediate Representations (IRs). 
%Since we have implemented \name for x86 architecture, it is possible to 
%To this end, we can extend the symbolic execution module and IR translation module to support other architectures.
%\mypara{Cross-architecture Support}

%For OLLVM obfuscation option of {FLA}, \name performed not as well as Asm2Vec~\cite{asm2vec}. 
%We also discussed the future work and limitations of \name.
%% For numbered reference style

%\bibitem{lamport94}
%  \textit{\LaTeX: a document preparation system},
%  Addison Wesley, Massachusetts,

%\end{thebibliography}

\bibliographystyle{apalike}
\bibliography{reference}
\end{document}